\documentclass{aa}
\usepackage[applemac]{inputenc}
\usepackage{natbib}
\usepackage{txfonts}
\usepackage{graphicx}
\usepackage{color}
\bibpunct{(}{)}{;}{a}{}{,}

\begin{document}

\title{Benchmarking spin-state chemistry in starless core models}
\author{O. Sipilä\inst{1,2},
	     P. Caselli\inst{1},
	     \and{J. Harju\inst{2,1}}
}
\institute{Max-Planck-Institute for Extraterrestrial Physics (MPE), Giessenbachstr. 1, D-85748 Garching, Germany\\
e-mail: \texttt{osipila@mpe.mpg.de}
\and{Department of Physics, PO Box 64, 00014 University of Helsinki, Finland}
}

\date{Received / Accepted}

\abstract
{}
{We aim to present simulated chemical abundance profiles for a variety of important species, with special attention given to spin-state chemistry, in order to provide reference results against which present and future models can be compared.}
{We employ gas-phase and gas-grain models to investigate chemical abundances in physical conditions corresponding to starless cores. To this end, we have developed new chemical reaction sets for both gas-phase and grain-surface chemistry, including the deuterated forms of species with up to six atoms and the spin-state chemistry of light ions and of the species involved in the ammonia and water formation networks. The physical model is kept simple in order to facilitate straightforward benchmarking of other models against the results of this paper.}
{We find that the ortho/para ratios of ammonia and water are similar in both gas-phase and gas-grain models, at late times in particular, implying that the ratios are determined by gas-phase processes. Furthermore, the ratios do not exhibit strong dependence on core density. We derive late-time ortho/para ratios of $\sim 0.5$ and $\sim 1.6$ for ammonia and water, respectively. We find that including or excluding deuterium in the calculations has little effect on the abundances of non-deuterated species and on the ortho/para ratios of ammonia and water, especially in gas-phase models where deuteration is naturally hindered owing to the presence of abundant heavy elements. Although we study a rather narrow temperature range ($10-20$\,K), we find strong temperature dependence in, e.g., deuteration and nitrogen chemistry. For example, the depletion timescale of ammonia is significantly reduced when the temperature is increased from 10 to 20\,K; this is because the increase in temperature translates into increased accretion rates, while the very high binding energy of ammonia prevents it from being desorbed at 20\,K.}
{}

\keywords{ISM: abundances -- ISM: clouds -- ISM: molecules -- astrochemistry}

\maketitle

\section{Introduction}

The ortho-to-para ratio of molecular hydrogen, $\rm H_2$, plays a large role in the development of deuterium chemistry at the high density and low temperature attributed to starless cores. The deuteration sequence begins with the exothermic reaction between $\rm H_3^+$ and HD:
\begin{equation}\label{eq:deutini}
\rm H_3^+ + HD \longleftrightarrow H_2D^+ + H_2 + 232\,K \, .
\end{equation}
The $\rm H_2D^+$ ion can then donate its deuteron to other abundant species, such as CO and $\rm N_2$, yielding $\rm DCO^+$ and $\rm N_2D^+$, respectively. However, spin-state effects complicate the reaction scheme. Chemical species with multiple protons, or deuterons, can exist in different spin configurations -- for example $\rm H_2$ has two distinct spin states where the nuclear spin wavefunction is either symmetric (ortho-$\rm H_2$; hereafter o$\rm H_2$) or antisymmetric (para-$\rm H_2$; hereafter p$\rm H_2$). The difference in energy between the ground states of these two spin states, $\sim 170$\,K \citep{Hugo09}, can be a very important energy reservoir at the low temperatures of starless cores. Indeed, reaction\,(\ref{eq:deutini}) can proceed relatively easily in the backward direction when the ortho forms of both $\rm H_2D^+$ and $\rm H_2$ are involved. Consequently, the ortho-to-para (hereafter o/p) ratio of $\rm H_2$ is an important parameter controlling deuterium chemistry \citep{Flower06a}. In addition to deuteration, spin-state effects play a large role in other areas of starless core chemistry as well. For example, the formation chain of ammonia depends critically on the $\rm N^+ + H_2$ reaction, which requires the presence of o$\rm H_2$ to proceed efficiently at low temperatures \citep{Dislaire12}.

Owing to its importance, spin-state chemistry is now widely adopted in numerical chemical models, and has been applied to all the stages in the star formation process, i.e., diffuse clouds \citep{Albertsson14b}, starless/prestellar cores (\citealt{WFP04}; \citealt{FPW04}; \citealt{Pagani09}; \citealt{Sipila10, Sipila13}; hereafter S13) and protostellar systems (\citealt{Taquet13}; \citealt{Taquet14}). However, spin-state chemistry is usually discussed in the context of specific problems adopting very different physical models, and comparison between the results of different works can be difficult. In this paper, we aim to remedy this by presenting relatively easily reproducible reference results pertaining to spin-state chemistry, in physical conditions appropriate to starless cores. To this end, we have developed a state-of-the-art spin-state chemical model that includes the spin states of light hydrogen-containing species ($\rm H_2$, $\rm H_3^+$...), and the spin states of the species taking part in the formation and destruction networks of ammonia and water. Our model also includes the deuterated forms of species with up to six atoms. To extend the usability of our results, we present results for both gas-phase and gas-grain models.

The paper is organized as follows. Sect.\,\ref{s:models} describes our physical and chemical models in detail. In Sect.\,\ref{s:results}, we present the results of our calculations. In Sect.\,\ref{s:discussion}, we discuss our results and in Sect.\,\ref{s:conclusions} we present our conclusions. Appendices \ref{appendix:latetime} to \ref{appendix:tunneling} include complementary discussion on our main results, and also present additional modeling results.

\section{Model}\label{s:models}

\subsection{Physical parameters}

In order to facilitate straightforward comparison of future modeling works against the results presented in this paper, we consider homogeneous models.  That is, we fix the values of all physical parameters but the density of the medium. The model parameters (introduced below) along with their assumed values are presented in Table~\ref{tab1}. In all calculations, we set the gas temperature $T_{\rm gas}$ equal to the dust temperature $T_{\rm dust}$. Although we restrict the temperature to $T = 10$\,K in the main text, we present reference results also for $T  = 15$\,K and $T = 20$\,K in Appendix\,\ref{appendix:difftemp} (these results are briefly discussed in Sect.\,\ref{ss:hightemp}).

The binding energies of the various species, corresponding to a water ice surface, are mainly taken from Garrod \& Herbst (\citeyear{Garrod06}; see also \citealt{Sipila12}). Table\,\ref{tab2} presents the binding energies of selected species. We assume that the binding energy of a deuterated species is equal to that of the corresponding undeuterated species. This approach has been previously adopted by, e.g., \citet{Cazaux10}, \citet{Taquet13} and S13. To cover typical density values associated with starless cores, we present results at four different densities ranging from $10^3$ to $10^6$\,cm$^{-3}$. 

\begin{table}
\caption{Adopted values of the various physical parameters (see text for definitions of the parameters).}
\centering
\begin{tabular}{c c}
\hline \hline 
Parameter & Value  \\ \hline
$T_{\rm gas} = T_{\rm dust}$ & 10\,K \\
$\zeta$ & $1.3\times10^{-17} \, \rm s^{-1}$ \\
$A_{\rm V}$ & 10 mag \\
$a_{\rm g}$ & $0.1 \, \rm \mu m$ \\
$\rho_{\rm g}$ & $3.0 \rm \, g \, cm^{-3}$ \\
$n_s$ & $1.5 \times 10^{15} \, \rm cm^{-2}$ \\
$E_{\rm d} / E_{\rm b}$ & 0.77 \\
$R_{\rm d}$ & 0.01 \\
\hline
\end{tabular}
\label{tab1}
\end{table}

\begin{table}
\caption{Binding energies (corresponding to a water ice surface) of selected species.}
\centering
\begin{tabular}{c c}
\hline \hline 
Species & Binding energy\,[K]  \\ \hline
$\rm H$ & 450 \\
$\rm H_2$ & 500 \\
$\rm C$ & 800 \\
$\rm N$ & 800 \\
$\rm N_2$ & 1000 \\
$\rm O_2$ & 1000 \\
$\rm CO$ & 1150 \\
$\rm O$ & 1390 \\
$\rm NH$ & 2378 \\
$\rm OH$ & 2850 \\
$\rm NH_2$ & 3956 \\
$\rm NH_3$ & 5534 \\
$\rm H_2O$ & 5700 \\
\hline
\end{tabular}
\label{tab2}
\end{table}

\subsection{Chemical code}

We use the gas-grain chemical code discussed in \citet{Sipila10}, \citet{Sipila12} and S13, which uses the rate equation approach to calculate chemical evolution in the gas phase and on grain surfaces. The reaction rate coefficient is, for the majority of gas-phase reactions, defined with the modified Arrhenius equation as
\begin{equation}
k = \alpha \left( T / 300\,{\rm K} \right)^{\beta} \exp \left(-\gamma / T \right) \, .
\end{equation}
However, there are some exceptions. For photodissociation reactions, the rate coefficient is defined as
\begin{equation}\label{eq:kphoto}
k_{\rm photo} = \alpha \exp \left(-\gamma_1 A_{\rm V} \right) \, ,
\end{equation}
where $A_{\rm V}$ is the visual extinction. For cosmic-ray-induced dissociation, the rate coefficient is defined as
\begin{equation}\label{eq:kcr}
k_{\rm CR} = \gamma_2 \zeta \, ,
\end{equation}
where $\zeta$ is the cosmic ray ionization rate. In Eqs.\,(\ref{eq:kphoto}) and (\ref{eq:kcr}), $\gamma_1$ and $\gamma_2$ are efficiency factors.

The abundance of dust grains is calculated as
\begin{equation}\label{eq:grainden}
X_{\rm g} = \frac{n_{\rm g}}{n_{\rm H}} = R_{\rm d} \frac{\mu_{\rm mol} \, m_{\rm H}}{\frac{4}{3} \pi a_{\rm g}^3 \, \rho_{\rm g}} \, ,
\end{equation}
where $\mu_{\rm mol}$ is the mean molecular weight per H atom (1.4); $\rho_{\rm g}$ is the grain material density; $R_{\rm d}$ is the dust-to-gas mass ratio. For gas-phase chemical reactions involving a grain and another reactant with different electric charge, the rate coefficient is multiplied by the ``J-factor'' taking into account increased reaction efficiency due to polarization \citep{Draine87, Pagani09, Sipila10}.

Gas-phase chemistry is linked with grain-surface chemistry through adsorption and desorption processes. The adsorption rate coefficient of species $i$ is given by
\begin{equation}
k_i^{\rm ads} = {\upsilon}_i \, S \sigma \, ,
\end{equation}
where ${\upsilon}_i = \sqrt{8 k_{\rm B} T_{\rm gas} / \pi m_i}$ is the thermal speed of species $i$ ($k_{\rm B}$ is the Boltzmann constant and $m_i$ is the mass of species $i$); $S$ is the sticking coefficient, set to unity for all species; $\sigma = \pi a_{\rm g}^2$ is the grain cross section assuming spherical grains with radius~$a_{\rm g}$. The adopted desorption mechanism is cosmic-ray induced desorption, with rate coefficient given by
\begin{equation}
k_i^{\rm des} = f_{70} \, k_i^{\rm TD} (70 \, {\rm K}) = f_{70} \, \nu_{0,i} \exp \left[ -E_{{\rm b},i} / 70\, {\rm K} \right] \, ,
\end{equation}
where $f_{70} = 3.16 \times 10^{-19}$ is an efficiency factor; $\nu_{0,i} = \sqrt{2 n_s k_{\rm B} E_{{\rm b},i} / \pi^2 m_i}$ is the characteristic vibration frequency of species $i$ \citep{HHL92, HH93}. In the above, $n_s$ and $E_{{\rm b},i}$ stand for the density of binding sites on the grain surface and the binding energy of species $i$ on the grain surface, respectively. For simplicity, we consider only cosmic-ray induced desorption in this work, and neglect alternative desorption mechanisms such as photodesorption or reactive desorption.

The rate coefficient for a grain-surface reaction between species $i$ and $j$ is given by
\begin{equation}\label{eq:ratecoeff}
k_{ij} = \alpha \, \kappa_{ij} \left( R_i^{\rm diff} + R_j^{\rm diff} \right) / n_{\rm g} \,
\end{equation}
where $\alpha$ is the branching ratio of the reaction; the efficiency factor $\kappa_{ij} = \exp \left( - E_{\rm a} / T_{\rm dust} \right)$ or unity for exothermic reactions with or without activation energy ($E_{\rm a}$), respectively; $n_{\rm g}$ is the number density of dust grains (see below). The (thermal) diffusion rate $R^{\rm diff}$ is given by
\begin{equation}\label{eq:diff}
R_i^{\rm diff} = {\nu_i \over N_s} \exp(-E_{{\rm d},i} / T_{\rm dust}) \, ,
\end{equation}
where $N_s = n_s \, 4\pi a_g^2$ is the number of binding sites on the grain and $E_{{\rm d},i}$ is the diffusion energy of species $i$. The diffusion energy is determined by assuming a constant value for the diffusion-to-binding-energy ratio $E_{\rm d} / E_{\rm b}$.

We assume that the gas is initially atomic with the exception of hydrogen and deuterium which are locked in $\rm H_2$ and HD, respectively (Table~\ref{tab3}). We used the same assumptions about the initial abundances as \citet{Semenov10} except for HD for which we adopted $n({\rm HD})/n({\rm H}) = 1.6 \times 10^{-5}$ based on the D/H ratio measured in the local ISM \citep{Linsky95}. The initial $\rm H_2$ ortho/para ratio is (arbitrarily) set to $1 \times 10^{-3}$.

Finally we note that we do not consider quantum tunneling on grain surfaces in the main body of the text. However, reference results including tunneling are presented in Appendix~\ref{appendix:tunneling}, where we also discuss the modifications to the gas-grain model required when tunneling is included.

\begin{table}
\caption{Initial chemical abundances with respect to $n_{\rm H}$, and the adopted initial $\rm H_2$ o/p ratio.}
\centering
\begin{tabular}{c c}
\hline \hline 
Species & Initial abundance  \\ \hline
$\rm H_2$ & 0.5 \\
$\rm He$ & $9.00\times10^{-2}$  \\
$\rm HD$ & $1.60\times10^{-5}$ \\
$\rm O$ & $2.56\times10^{-4}$ \\ 
$\rm C^+$ & $1.20\times10^{-4}$  \\
$\rm N$ & $7.60\times10^{-5}$ \\
$\rm S^+$ & $8.00\times10^{-8}$ \\
$\rm Si^+$ & $8.00\times10^{-9}$ \\
$\rm Na^+$ & $2.00\times10^{-9}$ \\
$\rm Mg^+$ & $7.00\times10^{-9}$ \\
$\rm Fe^+$ & $3.00\times10^{-9}$ \\
$\rm P^+$ & $2.00\times10^{-10}$ \\
$\rm Cl^+$ & $1.00\times10^{-9}$ \\
$\rm H_2\,(o/p)_{\rm ini}$ & $1.00\times10^{-3}$ \\
\hline
\end{tabular}
\label{tab3}
\end{table}

\subsection{Gas-phase and grain-surface reaction sets}

The chemical reaction sets for gas-phase and grain-surface reactions used in this work are reworked versions of the networks presented in S13; the present gas-phase network is based on {\tt osu\_01\_2009}\footnote{See {\tt  http://www.physics.ohio-state.edu/$\sim$eric/}} instead of the modified version of {\tt osu\_03\_2008} \citep{Semenov10} adopted in S13. In the present model, the spin-state chemistry description of S13 is expanded by adding the spin states of the species involved in the formation of water and ammonia. Also, the present model contains deuterated forms of species with up to 6 atoms, including observationally important species such as water, ammonia and methanol. We discuss the various updates and additions in detail in the following.

\subsubsection{The spin-state separation routine}\label{ss:sroutine}

In S13, a routine was developed to automatically spin-state separate any reaction involving light hydrogen-bearing species ($\rm H_2$, $\rm H_2^+$ and $\rm H_3^+$). In practice, the routine creates new reactions based on pre-determined branching ratios. In S13, the branching ratios for {\sl most reactions} were deduced using the method of \citet{Oka04}, in which representations of the nuclear spin wavefunctions of the various species are used to derive selection rules for reactive collisions. The resulting branching ratios correspond to pure nuclear spin statistical weights under the assumption that the nuclei are completely mixed in the reaction, and thus they are likely to be most applicable to highly exothermic reactions \citep{Oka04}. For example, the reaction
\begin{equation}\label{reac1}
{\rm H_3^+} + {\rm O} \mathop{\longrightarrow}\limits^k {\rm OH^+} + {\rm H_2},
\end{equation}
with rate coefficient $k$, separates into three reactions when the spin states of the species containing multiple protons are considered explicitly: 
\begin{eqnarray}\label{h3+sep}
{\rm H_3^+(o)} &+& {\rm O} \mathop{\longrightarrow}\limits^k {\rm OH^+} + {\rm H_2(o)} \nonumber \\
{\rm H_3^+(p)} &+& {\rm O} \mathop{\longrightarrow}\limits^{{1\over2}k} {\rm OH^+} + {\rm H_2(o)} \\
{\rm H_3^+(p)} &+& {\rm O} \mathop{\longrightarrow}\limits^{{1\over2}k} {\rm OH^+} + {\rm H_2(p)} \nonumber \, .
\end{eqnarray}
In S13, the branching ratio matrices for some reacting systems were derived by hand and applied in the o/p separation routine. However, for more complicated reactions involving many protons and/or multiple products, custom separation rules were applied (see Appendix~A in S13). The motivation for doing so was threefold. Firstly, Oka's method may not be applicable when the exothermicity of a given reaction is very low. Secondly, we did not want to follow the spin states of species heavier than $\rm H_3^+$ (with the exception of its deuterated forms), so in any reaction where such species (e.g., $\rm H_2S$) were present, we assumed that the species is in its para form. Thirdly, the branching ratio matrices were deduced by hand, which quickly becomes tedious when multiple protons and product species are present, even though the separation method itself is straightforward.

For this paper, we have developed a new version of the separation routine. In the new version, the branching ratios are calculated automatically using Oka's method so that we easily obtain the branching ratio matrix of any reaction. Consequently, {\sl every} species with multiple protons is separated into its ortho and para (and meta when applicable) states. However, in order to maintain relative simplicity of the reaction network, we only keep the spin states of the light hydrogen-bearing species, and of those species included in the formation networks of water and ammonia (see Sect.\,\ref{ss:newspins}).

When spin species other than those mentioned above are present in a reaction, we recombine the reactions over the ``unwanted'' spin species. For example, consider the reaction
\begin{equation}\label{reac_ex}
{\rm H_3^+} + {\rm CH_2} \mathop{\longrightarrow}\limits^{k_1} {\rm CH_3^+} + {\rm H_2}.
\end{equation}
When separated using spin selection rules, reaction~(\ref{reac_ex}) branches into the following set of reactions:
\begin{eqnarray}
{\rm oH_3^+} + {\rm oCH_2} &\mathop{\longrightarrow}\limits^{\frac{37}{60}k_1}& {\rm oCH_3^+} + {\rm oH_2} \nonumber \\
 					     &\mathop{\longrightarrow}\limits^{\frac{5}{60}k_1}& {\rm oCH_3^+} + {\rm pH_2} \nonumber \\
					     &\mathop{\longrightarrow}\limits^{\frac{14}{60}k_1}& {\rm pCH_3^+} + {\rm oH_2} \nonumber \\
					     &\mathop{\longrightarrow}\limits^{\frac{4}{60}k_1}& {\rm pCH_3^+} + {\rm pH_2} \nonumber \\
{\rm oH_3^+} + {\rm pCH_2} &\mathop{\longrightarrow}\limits^{\frac{1}{4}k_1}& {\rm oCH_3^+} + {\rm oH_2} \nonumber \\
 					     &\mathop{\longrightarrow}\limits^{\frac{1}{4}k_1}& {\rm oCH_3^+} + {\rm pH_2} \nonumber \\
					     &\mathop{\longrightarrow}\limits^{\frac{2}{4}k_1}& {\rm pCH_3^+} + {\rm oH_2} \nonumber \\
{\rm pH_3^+} + {\rm oCH_2} &\mathop{\longrightarrow}\limits^{\frac{14}{60}k_1}& {\rm oCH_3^+} + {\rm oH_2} \nonumber \\
 					     &\mathop{\longrightarrow}\limits^{\frac{10}{60}k_1}& {\rm oCH_3^+} + {\rm pH_2} \nonumber \\
					     &\mathop{\longrightarrow}\limits^{\frac{28}{60}k_1}& {\rm pCH_3^+} + {\rm oH_2} \nonumber \\
					     &\mathop{\longrightarrow}\limits^{\frac{8}{60}k_1}& {\rm pCH_3^+} + {\rm pH_2} \nonumber \\
{\rm pH_3^+} + {\rm pCH_2} &\mathop{\longrightarrow}\limits^{\frac{1}{5}k_1}& {\rm oCH_3^+} + {\rm oH_2} \nonumber \\
					    &\mathop{\longrightarrow}\limits^{\frac{2}{5}k_1}& {\rm pCH_3^+} + {\rm oH_2} \nonumber \\
					    &\mathop{\longrightarrow}\limits^{\frac{2}{5}k_1}& {\rm pCH_3^+} + {\rm pH_2} \nonumber \, .
\end{eqnarray}
To determine the branching ratios of the $\rm oH_3^+ + CH_2 \longrightarrow CH_3^+ + (o/p)H_2$ system, i.e., neglecting the spin states of $\rm CH_2$ and $\rm CH_3^+$, we sum over the $\rm oH_2$ and $\rm pH_2$ production pathways and average the rate coefficients on the number of spin modifications of $\rm CH_2$ (two). A similar treatment of the $\rm pH_3^+ + CH_2 \longrightarrow CH_3^+ + (o/p)H_2$ system finally results in four reactions:
\begin{eqnarray}
{\rm oH_3^+} + {\rm CH_2} &\mathop{\longrightarrow}\limits^{\frac{4}{5}k_1}& {\rm CH_3^+} + {\rm oH_2}  \label{eq13} \\
 					     &\mathop{\longrightarrow}\limits^{\frac{1}{5}k_1}& {\rm CH_3^+} + {\rm pH_2}  \\
{\rm pH_3^+} + {\rm CH_2} &\mathop{\longrightarrow}\limits^{\frac{13}{20}k_1}& {\rm CH_3^+} + {\rm oH_2} \\
 					     &\mathop{\longrightarrow}\limits^{\frac{7}{20}k_1}& {\rm CH_3^+} + {\rm pH_2} \label{eq16} \, .
\end{eqnarray}

The branching ratios in Eqs.\,(\ref{eq13}) to (\ref{eq16}) are based on the assumption that the ortho and para forms of $\rm CH_2$ are equally abundant. Using the statistical ortho/para $\rm CH_2$ ratio (3:1), the four branching ratios are 33/40, 7/40, 27/40 and 13/40, respectively, while assuming that $\rm CH_2$ is completely in the para state yields the branching ratios 3/4, 1/4, 3/5 and 2/5. If we assume that the reaction proceeds through a proton hop process instead of complete scrambling, we obtain the branching ratios 1, 0, 1/2 and 1/2 (\citealt{Oka04}, Table 8b). Evidently, there are differences in the outcome depending on the assumed ortho/para ratio of $\rm CH_2$ or the assumed reaction mechanism (which is not known for the vast majority of reactions). We note, however, that heavier molecules have a negligible effect on the $\rm H_2$ ortho/para ratio for example, and our assumptions seem justified in the present case where we only consider explicitly the spin states of a limited set of species.

It should be noted that the Oka method allows the formation of o$\rm H_2$ from p$\rm H_2$, simply because this is statistically possible. However, the formation of $\rm oH_2$ should be unlikely at low temperatures where the collisional energies are small compared with the energy difference $\Delta E/k =170$\,K between the ground states of $\rm oH_2$ and $\rm pH_2$, unless the reaction is sufficiently exothermic. In S13, we assumed that the required exothermicity is not reached in general, and hence the network favored p$\rm H_2$ formation. In this paper, we slightly relax the assumptions of S13, and apply the separation routine to all reactions other than charge-transfer reactions, where we assume that spin states are conserved. However, we add an activation energy of 170\,K to the $\gamma$ coefficient in those reactions where o$\rm H_2$ is created by reactants whose spin states are not explicitly considered (for example, $\rm CH_3^+ + CH_2 \longrightarrow C_2H_3^+ + H_2$; see also \citealt{Albertsson14b}). The new approach modifies the results of S13 only very slightly; see Sect.\,\ref{ss:comparison}. With these changes, the resultant network should be better applicable at higher temperatures (\citealt{Albertsson14}; see also \citealt{Albertsson14b}). The combination of nuclear spin statistical weights and activation energies for probably endoergic reactions approximates the method \citet{Hugo09} used to calculate the thermal rate coefficients for the $\rm H_3^+ + H_2$ isotopic system, but we note that a detailed state-to-state analysis such as that performed by \citet{Hugo09} is required to study endothermic reactions consistently.

We apply the same spin-state separation routine to grain-surface reactions. No rate coefficient corrections have been made to the surface reactions because in the base surface reaction set adopted here \citep{Semenov10}, the non-addition reactions that create $\rm H_2^{\ast}$ (the asterisk denotes a surface species), for example $\rm H^{\ast} + H_2CO^{\ast} \longrightarrow HCO^{\ast} + H_2^{\ast}$, already present high activation barriers. Of course, $\rm oH_2^{\ast}$ can be created by (cr-induced) photodissociations; we assume that these events are sufficiently energetic so that o$\rm H_2^{\ast}$ and p$\rm H_2^{\ast}$ are produced in the statistical ratio 3:1.

We note that the conservation of nuclear spin may not hold for grain-surface reactions as there is experimental evidence to the contrary \citep{Fushitani02, Hama13}. We discuss this issue briefly in Sect.\,\ref{ss:surfacespinsep}.

\subsubsection{Spin-state chemistry of ammonia and water}\label{ss:newspins}

In this work, we have included the spin-state chemistry of nitrogen-containing species involved in the ammonia formation chain, recently studied using the Oka method by \citeauthor{LeGal14} (\citeyear{LeGal14}; see also \citealt{Flower06b}; \citealt{Dislaire12}; \citealt{Faure13}; \citealt{Rist13}). We have checked that our o/p separation routine produces the same branching ratios as given in Tables B.1. and B.2. in \citet{LeGal14} (see also Sect.\,\ref{ss:othermod}.)

Because of the high exothermicities of the hydrogen addition reactions on grain surfaces creating $\rm NH_2^{\ast}$ and $\rm NH_3^{\ast}$ ($\sim$ 4.2\,eV and 4.7\,eV, respectively, as calculated from Eqs.\,(3) and (4) in \citealt{Allen77} and using data presented in Appendix~C of \citealt{Du12}), we assume that the high-temperature statistical branching ratios for the formation reactions,
\begin{eqnarray*}
\rm NH^{\ast} + H^{\ast} &\mathop{\longrightarrow}\limits^{3/4}& \rm oNH_2^{\ast} \\
                                      &\mathop{\longrightarrow}\limits^{1/4}& \rm pNH_2^{\ast} \\
\rm oNH_2^{\ast} + H^{\ast} &\mathop{\longrightarrow}\limits^{2/3}& \rm oNH_3^{\ast} \\
                                      &\mathop{\longrightarrow}\limits^{1/3}& \rm pNH_3^{\ast} \\
\rm pNH_2^{\ast} + H^{\ast} &\mathop{\longrightarrow}\limits^{1}& \rm pNH_3^{\ast} \, ,
\end{eqnarray*}
are valid. These branching ratios are, of course, given directly by the Oka method.

We have also included the spin-state chemistry of the species involved in the water formation network. The branching ratios for the most important reactions in the water network are presented in Appendix\,\ref{appendix:water}. Similarly to the nitrogen chemistry, we adopt statistical branching for water formation on grain surfaces:
\begin{eqnarray}
\rm OH^{\ast} + H^{\ast} &\mathop{\longrightarrow}\limits^{3/4}& \rm oH_2O^{\ast} \nonumber \\
                                      &\mathop{\longrightarrow}\limits^{1/4}& \rm pH_2O^{\ast} \, .
\end{eqnarray}

\subsubsection{Deuteration}\label{sss:deuteration}

In S13, a deuteration routine was applied to the OSU network, and species with up to four atoms were deuterated. In the present work, the deuteration routine is extended to handle species with up to six atoms, so that the deuterated forms of important species such as methanol and ammonia (whose formation depends on $\rm NH_4^+$) can be included. The deuteration routine is based on that presented by \citet{Rodgers96}, in which deuterons are substituted in place of protons in the various reactions and branching ratios are calculated assuming complete scrambling (see also S13).

The calculation of the nuclear spin-state branching ratios for multiply-deuterated species is more complicated than for undeuterated species. For species with three or more deuterium nuclei Oka's method, based on angular momentum algebra, is insufficient because there is no one-to-one correspondence between angular momentum and symmetry representations. The statistical branching ratios between the nuclear spin states of species with up to five hydrogen or deuterium nuclei can be obtained from Tables III and IV of \citet{Hugo09}. In the present paper, the spin-state chemistry in reactions between species containing H and/or D {\sl only} is included using these rules, and they are mostly available in the literature. Specifically, we include the spin-state chemistry of the $\rm H_3^+ + H_2$ reacting system from \citet{Hugo09}, complemented by data for other reactions concerning light hydrogen-and-deuterium-bearing species ($\rm H_3^+$, $\rm H_2D^+$, HD etc.) from \citet{WFP04} and \citet{FPW04}. The rate coefficients for dissociative recombination between $\rm H_3^+$ and its deuterated forms and electrons are taken from \citet{Pagani09}.

The nuclear spin chemistry of multiply-deuterated molecules containing heavy elements is not considered in the present model, with the exception that it is necessary to predict the ratio of o$\rm D_2$ and p$\rm D_2$ released in
reactions involving heavy species to achieve a properly closed reaction network. Furthermore, o$\rm D_2$ and p$\rm D_2$ produced by these reactions can be relevant to the chemistry of light molecules containing only H and/or D which are treated using the appropriate spin selection rules (see below).

The deuteration procedure used here directly propagates the hydrogen spin-state chemistry into the deuterium chemistry, for molecules with up to two deuterium nuclei. For hydrogenated species like $\rm CH_2$, we assumed that the ortho and para forms of $\rm CH_2$ are equally abundant (see Sect. 2.3.1). However, copying the reaction scheme from hydrogenated to doubly deuterated molecules is equivalent to following the correct nuclear spin selection rules for spin 1 systems with the assumption that o$\rm D_2$ and p$\rm D_2$ are instead present in the ratio 2:1 (i.e., the statistical value) in complexes with heavy nuclei, e.g., $\rm CD_2$.

As an example of the spin-state separation of doubly-deuterated species, consider the reaction
\begin{equation}
{\rm H^+} + {\rm CH_2} \mathop{\longrightarrow}\limits^{k_2} \rm CH^+ + H_2 \, .
\end{equation}
Since our program performs the spin-state separation of hydrogen species prior to deuteration, the above reaction is first divided into the following two branches:
\begin{eqnarray}\label{h2ex}
{\rm H^+} + {\rm CH_2} &\mathop{\longrightarrow}\limits^{}& {\rm CH^+} + {\rm oH_2} \; , 
                         \;k=\frac{2}{3} k_2 \exp(-170/T) \; , \nonumber \\
                      &\mathop{\longrightarrow}\limits^{}& {\rm CH^+} + {\rm pH_2} \; ,
                         \;k=\frac{1}{3}k_2 \; .
\end{eqnarray}
The branching ratio (2/3 o$\rm H_2$, 1/3 p$\rm H_2$) follows from the nuclear spin statistics under the implicit assumption that p$\rm CH_2$ and o$\rm CH_2$ have equal abundances, i.e. $[\rm pCH_2]=[\rm oCH_2]$. Deuterating both branches leads in our routine to the following reactions:
\begin{eqnarray*}
{\rm H^+} + {\rm CHD} &\mathop{\longrightarrow}\limits^{\frac{2}{9}k_2\exp(-170/T)}& \rm CD^+ + oH_2 \\
                                     &\mathop{\longrightarrow}\limits^{\frac{1}{9}k_2}& \rm CD^+ + pH_2 \\
                                     &\mathop{\longrightarrow}\limits^{\frac{6}{9}k_2}& \rm CH^+ + HD \\
{\rm H^+} + {\rm CD_2} &\mathop{\longrightarrow}\limits^{\frac{2}{9}k_2}& \rm CH^+ + oD_2 \\
				&\mathop{\longrightarrow}\limits^{\frac{1}{9}k_2\exp(-86/T)}& \rm CH^+ + pD_2 \\
                                      &\mathop{\longrightarrow}\limits^{\frac{6}{9}k_2}& \rm CD^+ + HD \\
{\rm D^+} + {\rm CH_2} &\mathop{\longrightarrow}\limits^{\frac{2}{9}k_2\exp(-170/T)}& \rm CD^+ + oH_2 \\
                                      &\mathop{\longrightarrow}\limits^{\frac{1}{9}k_2}& \rm CD^+ + pH_2 \\
                                      &\mathop{\longrightarrow}\limits^{\frac{6}{9}k_2}& \rm CH^+ + HD \\
{\rm D^+} + {\rm CHD} &\mathop{\longrightarrow}\limits^{\frac{2}{9}k_2}& \rm CH^+ + oD_2 \\
				&\mathop{\longrightarrow}\limits^{\frac{1}{9}k_2\exp(-86/T)}& \rm CH^+ + pD_2 \\
                                      &\mathop{\longrightarrow}\limits^{\frac{6}{9}k_2}& \rm CD^+ + HD \\
{\rm D^+} + {\rm CD_2} &\mathop{\longrightarrow}\limits^{\frac{2}{3}k_2}& \rm CD^+ + oD_2 \\
				&\mathop{\longrightarrow}\limits^{\frac{1}{3}k_2\exp(-86/T)}& \rm CD^+ + pD_2 \, .
\end{eqnarray*}
Reactions yielding p$\rm H_2$ or o$\rm D_2$ are assumed to have no activation energies, whereas those producing p$\rm D_2$ are assumed to have an activation energy of 86 K, corresponding to the energy separation between the ground states of p$\rm D_2$ and o$\rm D_2$ \citep{Hugo09}. As noted above, the branching ratios for the reactions producing $\rm D_2$ are correct under the implicit assumption that $[\rm oCD_2]:[\rm pCD_2] =$ 2:1. However, as in the case of $\rm H_2$, reactions involving heavy species have a negligible effect on the $\rm D_2$ o/p ratio. Note that for the $\rm H^+ + CHD$ and $\rm D^+ + CH_2$ reactions above, one would expect $\rm H_2$ formation in the o/p ratio 3:1, but in the example the ratio is instead 2:1 because the deuterated reactions adopt the branching ratios from the undeuterated parent reactions, in this case reactions (\ref{h2ex}), where the three-proton system leads to an $\rm H_2$ o/p ratio of 2:1.

We have verified through testing that the spin separation of doubly-deuterated (heavy) species is of little consequence to the results presented later in this paper, and therefore the separation method used here seems reasonable for the present work. We will discuss the spin separation of multiply-deuterated molecules in depth
in a future dedicated paper.

\subsubsection{Other updates with respect to S13}\label{ss:updates}

In addition to the updates to the o/p separation and deuteration routines described above, the rate coefficients of several reactions have been updated. First of all, we now adopt the {\tt osu\_01\_2009} reaction set instead of (modified) {\tt osu\_03\_2008}, so any updates of the base reaction set, including anion chemistry, are included in our new model. The modifications made to {\tt osu\_03\_2008} by \citet{Semenov10}, already included in S13, have been incorporated to the new model as well.

We have also included new rate coefficient data from the literature. We adopt the rate coefficients given in Tables 12 and 13 in \citet{Albertsson13} -- except for reactions involving hydrogen and/or deuterium {\sl only} for which we use other data as detailed above. We have also included selected data from \citet{LeGal14}, which we summarize in Table~\ref{tabupdates}. We refer the reader to \citet{Albertsson13} and \citet{LeGal14} for the original data references.

The impact of the rate coefficient updates is analyzed in Sect.\,\ref{ss:comparison}.

\section{Results}\label{s:results}

\begin{figure*}
\centering
\includegraphics[width=17cm]{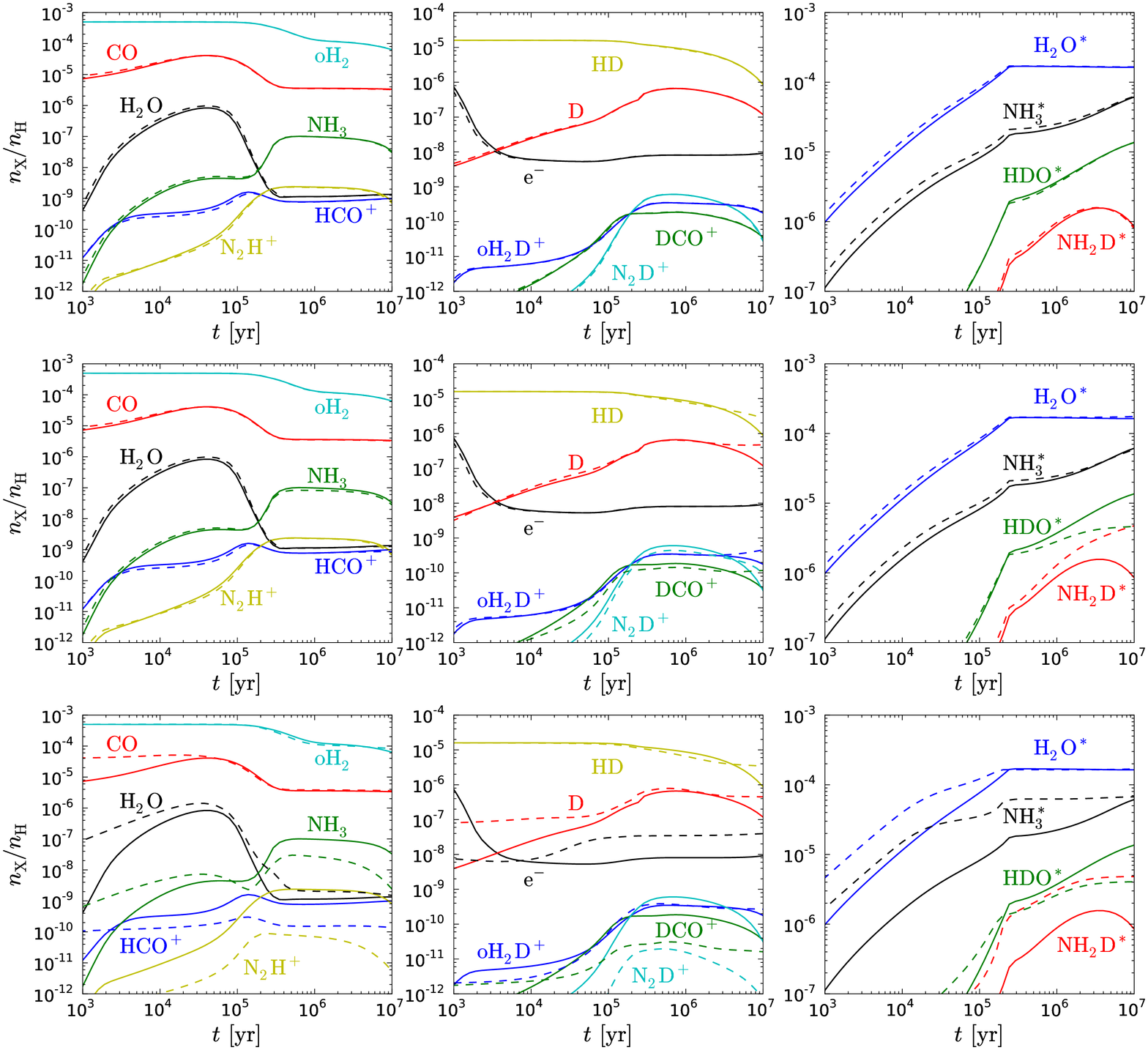}
\caption{{\sl Upper panels:} abundances of selected non-deuterated ({\bf left}), deuterated ({\bf middle}) and grain-surface (both non-deuterated and deuterated, {\bf right}) species as functions of time, as calculated with model T1 (solid lines) or model T2 (dashed lines). {\sl Middle panels:} as the upper row, but comparing models T1 (solid lines) and T3 (dashed lines). {\sl Lower panels:} as the other rows, but comparing models T1 (solid lines) and F1 (dashed lines). In all calculations, the density of the medium is set to $n_{\rm H} = 10^5 \, \rm cm^{-3}$. The abundances of $\rm H_2O$, $\rm NH_3$ and $\rm NH_2D$ represent sums over their respective ortho and para states.
}
\label{fig:rscomparison}
\end{figure*}

In this section, we present benchmarking results for spin-state chemistry for different values of density, using gas-phase and gas-grain models. All of the results presented below correspond to $T = 10$\,K. Results for $T = 15$\,K and $T = 20$\,K are presented in Appendix~\ref{appendix:difftemp}, while results without deuterium and including tunneling are presented in Appendices~\ref{appendix:nodeut} and \ref{appendix:tunneling}, respectively. The data for the various figures presented below are available from the authors upon request.

\subsection{Comparison of the new model against S13}\label{ss:comparison}

\begin{figure*}
\centering
\includegraphics[width=17cm]{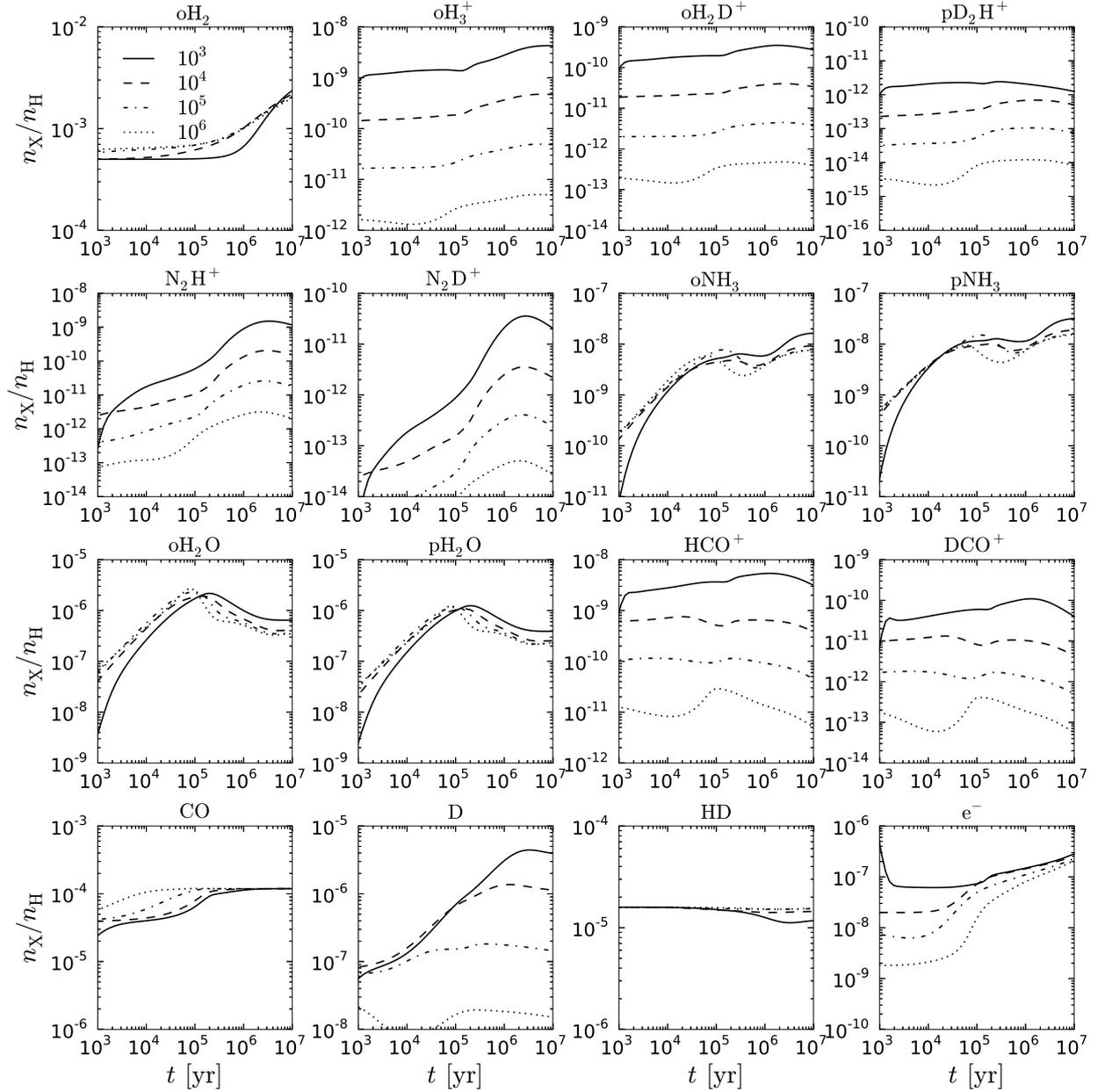}
\caption{Abundances of selected species calculated with model F2 (i.e., surface reactions and adsorption/desorption excluded) as functions of time at different densities, labeled in the upper left panel. The labels indicate densities of the medium; for example, $10^3$ stands for $n_{\rm H} = 10^3\,\rm cm^{-3}$.
}
\label{fig:gasphase10}
\end{figure*}

In our new model, there are two main sources of updates: 1) our new spin-state separation and deuteration routines and 2) the rate coefficient updates for various species from the literature and from the transition from {\tt osu\_03\_2008} to {\tt osu\_01\_2009}. Qualitatively, the new spin-state separation routine may affect somewhat the abundances of the various species because of the increased emphasis on o$\rm H_2$ in the new separation routine (albeit with a 170\,K barrier in most cases). However, the extension of deuteration from four to six atoms is not expected to have a large impact on the overall chemistry. In the following, we study the effect of the various updates, by successively introducing new updates into the S13 chemical model. The models introduced below are also described in Table \ref{tab4}.

\begin{table}
\caption{The models discussed in this work.}
\centering
\begin{tabular}{c c}
\hline \hline 
Model & Description  \\ \hline
T1 & the \citet{Sipila13} model \\
T2 & {\tt osu\_03\_2008} + new spin-state separation routine; \\
   & old deuteration routine; no rate coefficient updates \\
T3 & {\tt osu\_03\_2008} + new spin-state separation routine + \\
   & new deuteration routine; no rate coefficient updates \\
F1 & {\tt osu\_03\_2008} + new spin-state separation routine + \\
   & new deuteration routine + rate coefficient updates \\
   & (full model) \\
F2 & model F1 without gas-grain interaction (i.e., no \\
   & adsorption/desorption or surface reactions); \\
   & $\rm H_2$ formation from \citet{Kong14} \\
\hline
\end{tabular}
\label{tab4}
\end{table}

\begin{figure*}
\centering
\includegraphics[width=17cm]{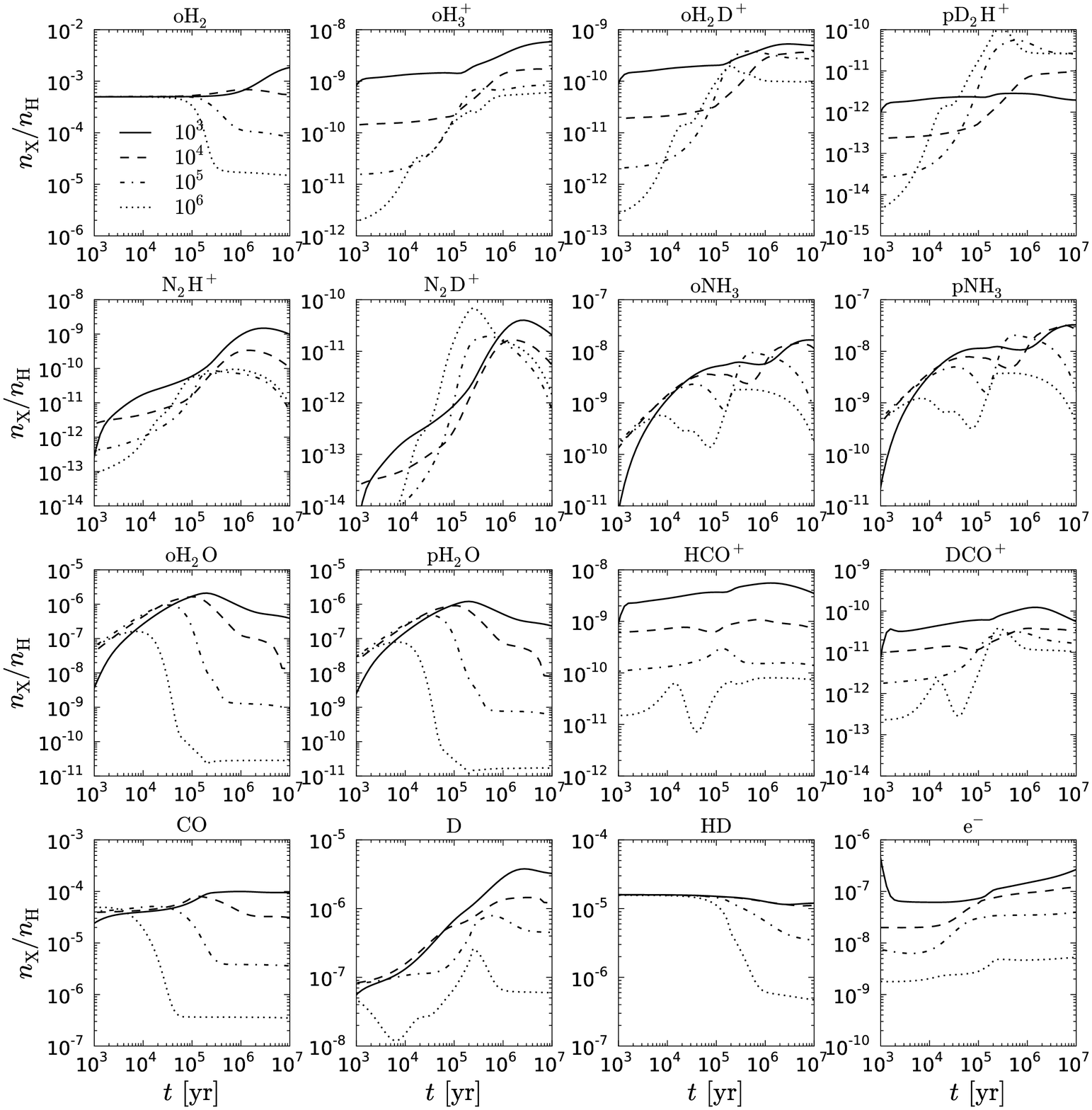}
\caption{As Fig.\,\ref{fig:gasphase10}, but calculated with model~F1. Note that the abundance scales are in some cases different than in Fig.\,\ref{fig:gasphase10}.
}
\label{fig:full10}
\end{figure*}

First, we constructed a test reaction set by spin-state separating the S13 base reaction set (modified {\tt osu\_03\_2008}) with the new routine presented here, but using the old S13 deuteration routine and including no new rate coefficient data (model~T2). The upper row in Fig.\,\ref{fig:rscomparison} plots the abundances of selected species as functions of time calculated using model~T2 (dashed lines) and using the old model of S13 (solid lines; model~T1). In these calculations, the density of the medium is set to $n_{\rm H} = 10^5 \, \rm cm^{-3}$, and the other physical parameters correspond to the values given in Table\,\ref{tab1}. In the right panel, the abundances of the respective ortho and para states of $\rm H_2O^{\ast}$, $\rm NH_3^{\ast}$ and $\rm NH_2D^{\ast}$ have been summed over. It is observed that the new spin-state separation routine modifies only slightly the abundances of non-deuterated species at early times both in the gas phase and on grain surfaces, while the abundances of deuterated species are virtually identical in both cases. Hence the new o$\rm H_2$-creating reactions with 170\,K barriers included here (as opposed to S13) are of little consequence to the chemistry as a whole (although at high temperatures the situation might be different).

The middle row in Fig.\,\ref{fig:rscomparison} plots the abundances of selected species calculated with a test reaction set constructed by applying the new spin-state separation and deuteration routines to the modified {\tt osu\_03\_2008} reaction set; as before, no new rate coefficient data is included (model~T3). It is observed that, overall, the extension of the deuteration scheme has little effect on the chemistry. However, deuteration is enhanced in the new model at long timescales because of less efficient HD depletion. Inspection of the reaction rates reveals that this effect is caused by the dissociation of ${\rm HDO^{\ast}}$, the main grain-surface deuterium carrier in our model, whose dissociation produced only $\rm OD^{\ast} + H^{\ast}$ in S13 owing to a bug in the deuteration routine, but here produces also $\rm OH^{\ast} + D^{\ast}$. Consequently, more $\rm D^{\ast}$ is available to form $\rm HD^{\ast}$, replenishing the gas-phase HD abundance.

In the bottom row of Fig.\,\ref{fig:rscomparison}, we plot again the abundances of selected species, now calculated with the full model presented in this paper, i.e., including the rate coefficient updates discussed in Sect.\,\ref{ss:updates} (model~F1). Evidently, the rate coefficient updates lead to significant differences with respect to the previous cases, both in the gas phase and on grain surfaces. For example, the CO formation timescale is much shorter with the updated rate coefficients, which is mainly due to numerous changes in electron DR rate coefficients. Accordingly, the electron abundance drops initially very rapidly in the new model. From an observational point of view, the most striking difference between the models is the abundance of $\rm N_2H^+$, which is over an order of magnitude lower at late times in the new model, resulting mainly from increased (factor 2-3) electron DR rate coefficients and also from changes in neutral-neutral chemistry, with respect to the OSU values. The changes is $\rm N_2H^+$ and $\rm HCO^+$ abundances propagate to their deuterated counterpart ions $\rm N_2D^+$ and $\rm DCO^+$. Also, the abundance of ammonia decreases by a factor of 2-3 at late times in the new model with respect to S13.

Surface chemistry is also affected by the rate coefficient updates. In the new model, atomic hydrogen is at early times produced mainly by neutral-neutral reactions between light hydrocarbons instead of electron DR reactions of various ions (owing to decreased electron abundance), and there is more atomic hydrogen available in the gas phase. The hydrogen is then adsorbed onto grain surfaces, leading to efficient production of, e.g., water and ammonia which form mainly by hydrogenation. We stress that these effects are caused by changes in the gas-phase chemistry, as none of the surface reaction rates have been updated. These results emphasize the importance of constraining the reaction rate coefficients, particularly for key reactions.

Our chemical model predicts gas-phase and grain-surface abundances in physical conditions attributed to starless cores, and our model results could in principle be used to interpret ice observations. However, direct observations of grain-mantle species toward starless cores are extremely hard owing to the very large extinctions. Thus, we would need to rely on, e.g., observations toward embedded Class~0 sources, assuming that the ice feature originates in the source envelope. A detailed investigation of modeled ice abundances versus observations is left for future work, and we focus our attention on gas-phase species in the present paper.

In what follows, all results have been produced with the new ortho/para separation and deuteration routines, and include the rate coefficient updates from the literature as described in Sect.\,\ref{ss:updates}.

\subsection{Gas-phase chemistry}

Figure \ref{fig:gasphase10} present the results of gas-phase modeling at different densities. Here, we have not considered adsorption and/or desorption of any species, and the formation of $\rm H_2$, HD and $\rm D_2$ on grain surfaces is included in the form of gas-phase reactions as in the model of \citet{Kong14}.

Ion abundances decrease significantly with increasing density, which is a consequence of the increased electron DR rates at high density, and of the high abundances of heavy neutral species owing to the absence of depletion. The abundances of heavy ions follow largely that of $\rm H_3^+$ which is primarily destroyed by reactions between abundant neutral species like CO and $\rm N_2$, and electrons. One example of this tendency is $\rm HCO^+$, which is the most important reaction partner of free electrons at late times both at low and high densities. This ion is formed in a reaction between $\rm H_3^+$ and CO and destroyed mainly in electron recombination which returns CO. At high densities, the enhanced production of free electrons by the cosmic ray ionization of $\rm H_2$ leads to an increase in the electron DR rates (at constant temperature), and the $\rm HCO^+$ abundance decreases accordingly. $\rm N_2H^+$ is locked in a similar cycle ($\rm H_3^+ + N_2 \longrightarrow N_2H^+$; $\rm N_2H^+ + e^- \longrightarrow N_2 + H$), dependent on the abundance of $\rm H_3^+$. Also the $\rm N_2H^+$ abundance decreases clearly when the density increases.

Deuteration is also suppressed in the absence of depletion, because $\rm H_3^+$ reacts preferentially with the abundant heavy species rather than with HD.

\subsection{Gas-grain chemistry}

Figure \ref{fig:full10} presents the results of gas-grain modeling at different densities, adopting the full model (F1) described above. Evidently, the inclusion of depletion significantly decreases the abundances of carbon and oxygen-containing species with respect to the gas-phase model. The $\rm H_3^+$ abundance is much higher at late times in the full model with respect to the gas-phase model because of the depletion of its main reaction partners (e.g., CO) onto grain surfaces. The chemistry of nitrogen-containing species is different than that of, e.g., carbon and oxygen because nitrogen chemistry depends on slow neutral-neutral reactions. On the other hand, the binding energy of (atomic and molecular) nitrogen is relatively low (Table \ref{tab2}), so nitrogen chemistry can still occur efficiently after other heavy elements have been depleted onto grain surfaces. Consequently, an appreciable abundance of $\rm N_2$, the precursor molecule of both $\rm NH_3$ and $\rm N_2H^+$ \citep[see, e.g.,][]{Fontani12}, can be present in the gas at late times. This is demonstrated by two effects. Firstly, the abundance of $\rm NH_3$ is less than an order of magnitude lower in the full model than in the gas-phase model at high density, even though depletion tends to decrease its abundance. Secondly, the abundance of $\rm N_2H^+$, dependent on $\rm H_3^+$, is orders of magnitude {\sl higher} with respect to the gas-phase model even at high density (we assume that ions do not deplete onto grain surfaces, but can recombine with negatively charged grains).

Deuteration increases strongly when gas-grain interaction is taken into account. However, a local peak in deuteration is observed around the time that HD starts deplete, in line with the results of S13. HD depletion is somewhat less severe in the present model than in S13 because we do not consider quantum tunneling. In agreement with previous results in the literature, our results clearly show that depletion is needed to produce observable amounts of, e.g., o$\rm H_2D^+$ and $\rm N_2H^+$ at high density.

Figures \ref{fig:gasphase10} and \ref{fig:full10} indicate that the o/p ratios of ammonia and water depend only slightly on density, both in the gas-phase and gas-grain models. We discuss this issue further below.

\section{Discussion}\label{s:discussion}

\subsection{The o/p ratios of water and ammonia}\label{amwatop}

\begin{figure}
\centering
\includegraphics[width=\columnwidth]{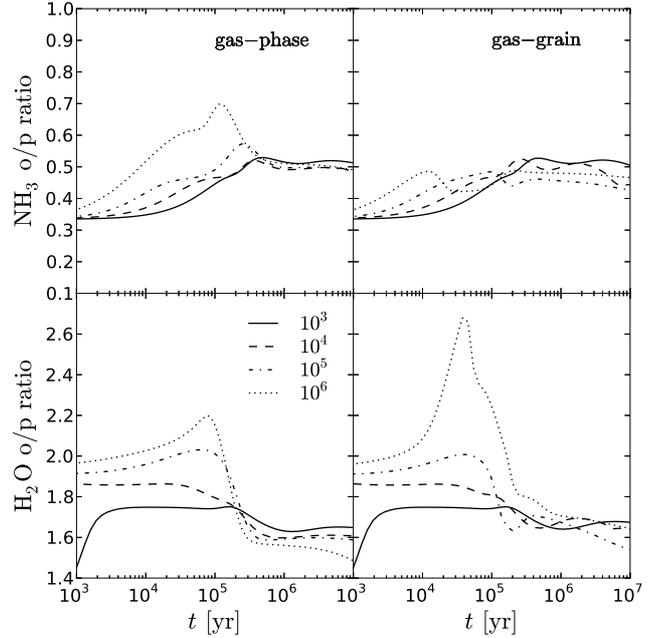}
\caption{o/p ratios of $\rm NH_3$ ({\sl upper panels}) and $\rm H_2O$ ({\sl lower panels}) as functions of time at different densities, labeled in the upper left panel. The left panels correspond to the gas-phase model (F2), while the right panels correspond to the gas-grain model (F1).
}
\label{fig:opratios}
\end{figure}

Figure \ref{fig:opratios} plots the o/p ratios of $\rm NH_3$ and $\rm H_2O$ at different densities according to gas-phase and gas-grain models. The o/p ratio of ammonia is very similar in both models up to $n_{\rm H} = 10^5 \, \rm cm^{-3}$. The ratio is larger at $\sim 10^4 - 10^5$ years at $n_{\rm H} = 10^6 \, \rm cm^{-3}$ in the gas-phase model, in which one of the main production pathways for $\rm NH_4^+$ at $t \sim 10^5$\,yr is the $\rm H_3O^+ + NH_3 \longrightarrow NH_4^+ + H_2O$ system which forms all three spin variants of $\rm NH_4^+$. However, m$\rm NH_4^+$ for example only dissociates to o$\rm NH_3$ \citep[see Table~2 in][]{LeGal14}. The detailed behavior of the spin-state chemistry is different in the gas-grain model, where the reaction pathways with $\rm H_3O^+$ are practically absent at $t \gtrsim 10^5$\,yr because of water depletion. Nevertheless our results indicate that, at late times, o/p-$\rm NH_3$ is $\sim 0.4 - 0.5$ regardless of the considered model or assumed density.

The o/p ratio of $\rm H_2O$ behaves similarly to that of $\rm NH_3$ in the sense that the largest differences between the gas-phase and gas-grain models are observed at $n_{\rm H} = 10^6 \, \rm cm^{-3}$. The o/p ratio of $\rm H_2O$ is tied to that of $\rm H_3O^+$, which evolves differently in the gas-grain model owing to multiple spin-state details ($\rm H_3O^+$ and $\rm H_2O$ are connected mainly through $\rm H_2O^+ + H_2 \longrightarrow H_3O^+ + H$ and the $\rm H_3^+ + H_2O \longrightarrow H_3O^+ + H_2$ system). The o/p-$\rm H_2O$ ratios at different densities and in different models are, again, very close to each other at late times where we derive o/p-$\rm H_2O \sim 1.5 - 1.7$.

Unlike the o/p ratio of $\rm H_2$ which is strongly influenced by grain-surface chemistry, the o/p ratios of $\rm H_2O$ and $\rm NH_3$ are determined nearly completely by gas-phase chemistry because of their high binding energies (see Table~\ref{tab2}). The fact that the o/p ratios of the respective species are very similar in both gas-phase and gas-grain models supports this interpretation (Fig.\,\ref{fig:opratios}). The late-time values of the o/p ratios can be justified by a rate coefficient analysis of the main formation and destruction paths of ammonia and water; such an analysis is presented in Appendix~\ref{appendix:latetime}.

We note that the late-time $\rm H_2O$ o/p ratios derived here are much lower than those typically observed in the ISM (\citealt{vanDishoeck13}; see also the discussion in \citealt{Keto14}). The disagreement is probably due to the fact that observations toward dark clouds with intermediate to high density are missing. To investigate this issue, we ran test calculations with our gas-grain model at conditions simulating translucent clouds ($n_{\rm H} = \rm 10^2\,cm^{-3}$ and $T = 10\,\rm K$) with two values of $A_{\rm V}$, 10\,mag and 1\,mag. We found that at $A_{\rm V} = 10\,\rm mag$, the $\rm H_2O$ o/p ratio is $\sim 1.6$, but at $A_{\rm V} = 1\,\rm mag$ the ratio is $\sim 3$, suggesting that photochemistry plays a large role in determining the o/p ratio in regions with low visual extinction. This result underlines the need for accurate physical models when interpreting observations.

Ammonia is a useful probe of the gas temperature in dark clouds \citep{Tafalla04, Juvela12}, and simulated $\rm NH_3$ line emission profiles can be used to constrain the kinetic temperature. However, a proper comparison of simulated line emission against a detection of, e.g., p$\rm NH_3$ requires an estimate of the $\rm NH_3$ o/p ratio. Given that the $\rm NH_3$ o/p ratio changes very little with density in our models, we deduce that o/p-$\rm NH_3 \sim 0.4-0.5$ is a good conservative estimate of the ratio in dark clouds. This result is consistent with observations by \citet{Persson12} and the models of \citet{LeGal14} (see also Sect.\,\ref{ss:othermod}).

Finally, we stress that we have not carried out a complete parameter-space study of the ammonia and water o/p ratios, and that such a study should be taken on in the future.

\subsection{Comparison against \citet{LeGal14}}\label{ss:othermod}

Recently, \citet{LeGal14} presented gas-phase modeling results of nitrogen chemistry in dark clouds. They included in their model the spin-state chemistry pertaining to the ammonia formation network, as we do here. We have run some test calculations to compare our results against those of \citet{LeGal14}, adopting the same physical conditions and initial chemical abundances for He, C, N and Fe (we choose $\rm [C]/[O] = 0.8$ and $[\rm S] = 8 \times 10^{-8}$). \citet{LeGal14} do not explicitly state the initial $\rm H_2$ o/p ratio; we assumed a value of~3. 

Table~\ref{tab5} summarizes the results of these tests. In our model, there are many cosmic-ray-induced and photodissociation reactions with long timescales, which prevent the system from reaching a true steady-state; the values given in the table correspond to late-time chemical evolution ($t = 2.0 \times 10^7$\,yr), after which temporal changes in chemical abundances are generally small. The Le Gal et al. steady-state values have been read off their Figs.~4~and~5. Evidently, our model gives lower $\rm NH$ and $\rm NH_2$ abundances and a higher $\rm NH_3$ abundance than the Le Gal et al. model, but the ortho/para ratios of both $\rm NH_2$ and $\rm NH_3$ are similar in both models.

We also tested an initial $\rm [C]/[O]$ ratio of 0.3, and found that in this case the late-time $\rm NH$ and $\rm NH_2$ abundances are lower by one and two orders of magnitude, respectively, than those predicted by \citet{LeGal14}. However, the $\rm NH_3$ abundance and the $\rm NH_2$ and $\rm NH_3$ o/p ratios are again similar to the \citet{LeGal14} model.

Our test results indicate that there are minor differences in fractional abundances between our model and that of \citet{LeGal14}. The two models predict similar o/p ratios for $\rm NH_2$ and $\rm NH_3$, which is expected since we adopt the same spin-state chemical description for these species as \citet{LeGal14}.

\begin{table}
\caption{Abundances and abundance ratios predicted by our gas-phase model (F2) and that of \citet{LeGal14}.}
\centering
\begin{tabular}{c c c}
\hline \hline 
Parameter & Our model & \citet{LeGal14}  \\ \hline
$\rm \left[ NH \right]$ & $5.9 \times 10^{-9}$ & $1.1 \times 10^{-8}$ \\
$\rm \left[ NH_2 \right]$ & $9.5 \times 10^{-10}$ & $3.5 \times 10^{-9}$ \\
$\rm \left[ NH_3 \right]$ & $3.2 \times 10^{-8}$ & $1.4 \times 10^{-8}$ \\
$\rm \left[ NH_2 \right] / \left[ NH \right]$ & $0.16$ & $0.3$ \\
$\rm \left[ NH_3 \right] / \left[ NH \right]$ & $5.4$ & $1.3$ \\
$\rm \left[ oNH_2 \right] / \left[ pNH_2 \right]$ & $1.8$ & $2.5$ \\
$\rm \left[ oNH_3 \right] / \left[ pNH_3 \right]$ & $0.4$ & $0.6$ \\
\hline
\end{tabular}
\label{tab5}
\end{table}

\subsection{$\rm H_2$ formation in gas-phase models}\label{ss:h2form}

\begin{figure}
\centering
\includegraphics[width=\columnwidth]{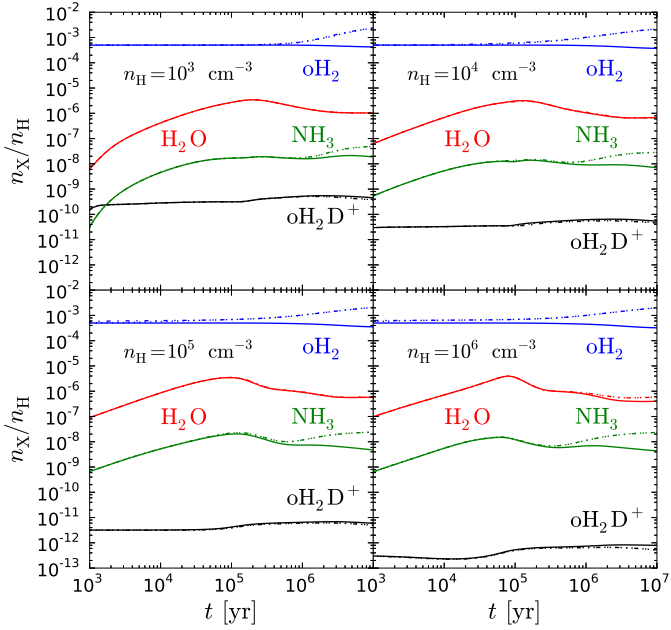}
\caption{Abundances of selected species as functions of time at different densities (labeled in the figure), in a gas-phase model at $T = 10$\,K, using different approaches to $\rm H_2$ formation. {\sl Dash-dotted lines} correspond to the gas-phase reactions of \citet{Kong14}. {\sl Solid lines} correspond to grain-surface $\rm H_2$ formation assuming only H and D adsorption and with the binding energies of H and $\rm H_2$ and their deuterated isotopologs set to 100\,K. {\sl Dotted lines} (superimposed on the dash-dotted lines) correspond to grain-surface $\rm H_2$ formation assuming only H and D adsorption and with the binding energies adopted in the main body of this work. The abundances of ammonia, water and $\rm H_2D^+$ represent sums over the abundances of their respective ortho and para states.
}
\label{fig:mrtest}
\end{figure}

In the gas-phase results presented above, the formation of $\rm H_2$ (and its isotopologs HD and $\rm D_2$) is included in the form of gas-phase reactions \citep[see][]{Kong14}. From Fig.\,\ref{fig:gasphase10} it is evident that this approach leads to, e.g., an increase of the $\rm H_2$ o/p ratio at late times. Another approach to including the formation of $\rm H_2$ and its isotopologs in a gas-phase model is to invoke grain-surface formation processes, but letting only atomic H and D to be adsorbed onto the grain surfaces.

We have run test calculations at $T = 10$\,K comparing the two approaches. The results of these calculations are presented in Fig.\,\ref{fig:mrtest}, where the abundances of selected species as functions of time are plotted at different densities, using different approaches to $\rm H_2$ formation. Evidently, the \citet{Kong14} approach is practically equivalent to the grain-surface formation approach when ``normal'' binding energies (450\,K for H and D) are used (the dotted and dash-dotted lines are practically superimposed). In the gas-grain model with very low H binding energy (100\,K), H desorbs very fast, leading to inefficient $\rm H_2$ production on the grain surfaces regardless of the density. Increasing the binding energy increases the average time spent by H atoms on the grains and thus leads to more efficient $\rm H_2$ production.

The choice of the $\rm H_2$ formation efficiency has little effect on the abundance of $\rm H_2D^+$ because deuterium chemistry is suppressed owing to the lack of depletion, and on the abundance of water because its formation pathway proceeds through exothermic reactions. However, the abundance of ammonia is slightly affected at late times because its formation starts with the $\rm N^+ + H_2$ reaction which is strongly endothermic when p$\rm H_2$ is involved \citep{LeGal14}.

We have checked that the o/p ratio of ammonia is hardly affected by the choice of the $\rm H_2$ formation efficiency, despite its effect on the total $\rm NH_3$ abundance. This is because the o/p ratio is mainly determined by reactions further up in the ammonia formation chain (see Appendix\,\ref{appendix:latetime}). The o/p ratio of water is not affected because it is not sensitive to the $\rm H_2$ o/p ratio. However, the o/p ratio of $\rm H_2D^+$ is affected because it is mainly determined by the $\rm H_2$ o/p ratio \citep{Pagani92, Gerlich02, Hugo07, Bruenken14}; the late-time $\rm H_2D^+$ o/p ratio changes by a factor of $\lesssim 2$ depending on the choice of binding energies.

\subsection{Conservation of spin in grain-surface reactions}\label{ss:surfacespinsep}

In this work, we have assumed that high-temperature statistical branching ratios are applicable to both gas-phase and grain-surface reactions. However, it has been previously suggested that different selection rules may apply for grain-surface reactions \citep{Fushitani02} and, in the case of $\rm H_2$ formation, that $\rm H_2$ ortho/para conversion may occur on the surface subsequent to $\rm H_2$ formation \citep{Watanabe10, Hama13}.

\begin{figure}
\centering
\includegraphics[width=\columnwidth]{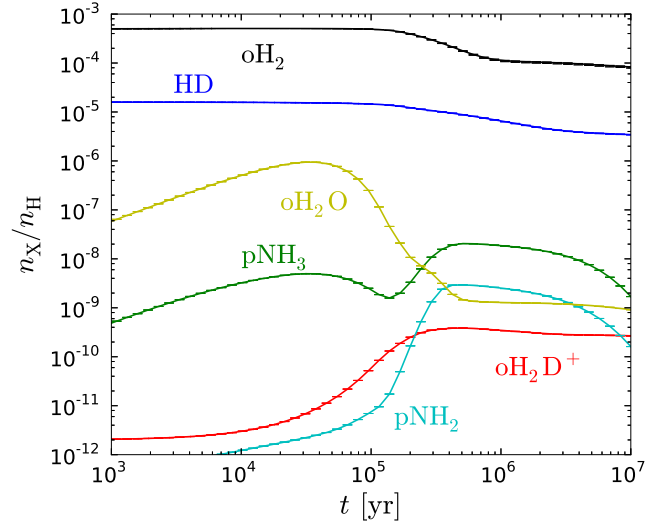}
\caption{Abundances of selected species as functions of time in a gas-grain model at $T = 10$\,K and $n_{\rm H} = 10^5\,\rm cm^{-3}$. The solid lines correspond to model~F1, while the error bars correspond to test models in which grain-surface reactions can only create ortho species (upper limits) or para species (lower limits).
}
\label{fig:surfacetest}
\end{figure}

To constrain the possible effect of uncertainties in grain-surface spin chemistry on our results, we constructed two test models in which grain-surface reactions can create either only para, or only ortho states -- with the exception of $\rm H_2$ and $\rm D_2$ for which we assume the high-temperature statistical formation ratios (i.e., both ortho and para are created). The result of this test is shown in Fig.\,\ref{fig:surfacetest}. Evidently, the effect of reactions other than the fundamental $\rm H^{\ast} + H^{\ast}$ reaction (and of its deuterated counterparts) on the gas-phase spin-state chemistry is negligible, as both the upper and lower limits correspond almost exactly to our normal full model.

We have carried out this test for all the medium densities considered in this paper ($n_{\rm H} = 10^3$ to $10^6\, \rm cm^{-3}$; with $T = 10\,\rm K$), and we conclude that the possible grain-surface branching ratio uncertainties related to spin-state chemistry do not modify the results of this paper. However, we note that \citet{Watanabe10} discussed $\rm H_2$ ortho/para conversion {\sl subsequent to} $\rm H_2$ formation; such conversion processes might translate to modifications in the gas-phase chemistry as the $\rm H_2$ molecules desorb. We currently have no means to test this, i.e., no data to form a rate coefficient. This issue should be investigated in the future.

\subsection{Results at higher temperatures}\label{ss:hightemp}

In Appendix~\ref{appendix:difftemp}, we present modeling results analogous to those shown in Figs.~\ref{fig:gasphase10}~to~\ref{fig:opratios}, but calculated assuming either $T_{\rm gas} = T_{\rm dust} = 15$\,K or $T_{\rm gas} = T_{\rm dust} = 20$\,K. No dramatic differences between the temperatures are seen in the gas-phase models, reflecting the weak temperature dependence of ion-molecule chemistry.

However, large differences between the temperatures are present in the gas-grain models. For example, nitrogen chemistry is much faster at high density at $T = 20$\,K than at $T = 10$\,K. The higher temperature increases the adsorption rates, but on the other hand the very high binding energy of ammonia for example means that practically no desorption occurs even at $T = 20$\,K. Consequently, ammonia and $\rm N_2H^+$ deplete almost totally from the gas phase in a timescale of $\sim 10^6$~years at $n_{\rm H} = 10^6\,\rm cm^{-3}$. Depletion occurs also at $T = 10$\,K, but in a much longer timescale (Fig.\,\ref{fig:full10}). We note that in the case of $\rm N_2H^+$, the ``depletion" is caused by the freeze-out of $\rm N_2$ onto grain surfaces, as we assume that molecular ions (e.g., $\rm N_2H^+$) dissociatively recombine with negatively charged grains and the products (in this case $\rm N_2 + H$) are immediately returned to the gas phase.

Deuteration is also strongly affected by changes in temperature. For example, the (late-time) deuteration degree of $\rm H_3^+$ is much lower at $T = 20$\,K than at $T = 10$\,K because the higher temperature causes 1) the deuteration reactions ($\rm H_3^+ + HD \longrightarrow H_2D^+ + H_2$ etc.) to proceed more efficiently in the backward direction and 2) less severe depletion of other important reaction partners of $\rm H_2D^+$ such as CO and $\rm N_2$, observed also as enchanced deuterations of $\rm HCO^+$ and $\rm N_2H^+$. Consequently, very little HD depletion is present at $T = 20$\,K regardless of density.

Although not readily apparent in the figures included here, we find a {\sl local minimum} in deuteration at around 10\,K if we study the abundances of deuterated species as a function of temperature but at fixed density. The temperature dependence is smoothed out if tunneling is included in the model, which also affects the deuteration degrees. This phenomenon might provide a means to investigate if tunneling is indeed effective on grain surfaces by studying the temperature dependence of the deuterium fractionation. This will be the subject of a future paper.

\section{Conclusions}\label{s:conclusions}

We have developed new chemical reaction sets for gas-phase and grain-surface chemistry, using the {\tt osu\_01\_2009} reaction set as a template. The new sets presented here are evolved versions of those presented in S13; the present model includes deuteration of species with up to six atoms (so that we can include the formation pathways of, e.g., ammonia and methanol), and the spin-state chemistry for the species involved in the water and ammonia formation networks.

We chose a simplified physical model, carrying out chemical calculations with a pseudo-time-dependent chemical code at different densities and temperatures (but with otherwise fixed physical parameters) in order to facilitate straightforward comparison of future modeling results against those presented in this paper. Calculations were performed with gas-phase and gas-grain models. Special attention was given to the ortho/para ratios of water and ammonia, and to the effect of either including or excluding deuterium in the chemical model.

We find that the o/p ratios of water and ammonia are $\sim 1.6$ and $\sim 0.5$, respectively, for $t > 10^5$\,yr irrespective of density or model (gas-phase or gas-grain) used. The o/p ratio of water is clearly lower than the value ($\sim 3$) observed toward translucent clouds, but we find that this apparent disagreement disappears if we consider low values of visual extinction (as opposed to $A_{\rm V} = 10$ assumed elsewhere in the present paper).

At early times, the o/p ratios vary with density, but little variation is seen between gas-phase and gas-grain models, implying that the o/p ratios are determined by gas-phase processes. Our results also show that, in the range of physical parameters considered, excluding deuterium from the model has only a marginal effect on the abundances of non-deuterated species. 

We find that an increase of the temperature from 10 to 20\,K generally decreases the depletion of heavy molecules onto grain surfaces, and this decreases also the deuteration degree. However, nitrogen chemistry proceeds differently; ammonia and $\rm N_2H^+$ deplete more strongly at $T = 20$\,K than at $T = 10$\,K.

The physical model of the present paper was deliberately kept simple in order to facilitate straightforward benchmarking of other models against the results presented here. However, we note that while the o/p ratio of ammonia did not show density dependence in our calculations, a more complete parameter-space study of ammonia (and water) spin-state chemistry, in the context of a gas-grain model, is called for.

\begin{acknowledgements}

We thank the anonymous referee for a thorough report and Malcolm Walmsley for helpful comments which improved the paper. O.S. acknowledges financial support of the Academy of Finland grant 250741, and of the Department of Physics of the University of Helsinki. P.C. acknowledges financial support of the European Research Council (ERC; project PALs 320620).

\end{acknowledgements}

\bibliographystyle{aa}
\bibliography{refs.bib}

\onecolumn
\twocolumn

\appendix

\newpage

\def\amm{\rm NH_3}
\def\pamm{\rm pNH_3}
\def\oamm{\rm oNH_3}

\def\nhfour{\rm NH_4^+}
\def\pnhfour{\rm pNH_4^+}
\def\onhfour{\rm oNH_4^+}
\def\mnhfour{\rm mNH_4^+}

\def\ammplus{\rm NH_3^+}
\def\pammplus{\rm pNH_3^+}
\def\oammplus{\rm oNH_3^+}

\def\el{\rm e^-}

\def\htwo{\rm H_2}
\def\phtwo{\rm pH_2}
\def\ohtwo{\rm oH_2}

\def\hthree{\rm H_3^+}
\def\phthree{\rm pH_3^+}
\def\ohthree{\rm oH_3^+}

\def\hplus{\rm H^+}

\def\htwoo{\rm H_2O}
\def\phtwoo{\rm pH_2O}
\def\ohtwoo{\rm oH_2O}

\def\hthreeo{\rm H_3O^+}
\def\phthreeo{\rm pH_3O^+}
\def\ohthreeo{\rm oH_3O^+}

\def\htwooplus{\rm H_2O^+}
\def\phtwooplus{\rm pH_2O^+}
\def\ohtwooplus{\rm oH_2O^+}

\def\ohplus{\rm OH^+}

\section{On the late-time ortho/para ratios of ammonia and water}\label{appendix:latetime}

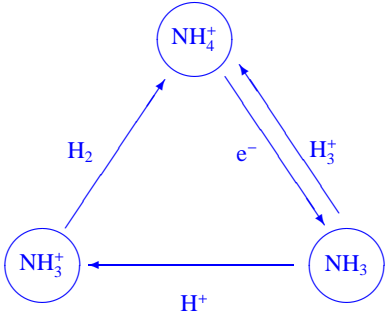
\begin{figure}[hbt]
   \centering
   \unitlength=1.0mm
  \begin{picture}(60,45)(0,0)
\color{blue} 

\put(10,10){\circle{10}}
\put(10,10){\makebox(0,0){$\ammplus$}}

\put(30,40){\circle{10}}
\put(30,40){\makebox(0,0){$\nhfour$}}

\put(50,10){\circle{10}}
\put(50,10){\makebox(0,0){$\amm$}}

\put(49,17){\vector(-2,3){13.1}}
\put(47,25){\makebox(0,0){$\hthree$}}

\put(34,35){\vector(2,-3){13.1}}
\put(37,25){\makebox(0,0){$\el$}}

\put(13,15){\vector(2,3){13.1}}
\put(15,25){\makebox(0,0){$\htwo$}}

\put(43,10){\vector(-1,0){27}}
\put(30,5){\makebox(0,0){$\hplus$}}

\end{picture}

  \caption{Ammonia cycle at late times of chemical evolution.}
      \label{figure:ammonia_cycle}
   \end{figure}

In the interstellar medium, ammonia production is dominated by the electron recombination of the ammonium ion, $\nhfour$, and its destruction occurs mainly through a charge transfer reaction with $\hplus$. At late times of chemical evolution, the proton transfer reaction with $\hthree$ returns ammonia back to $\nhfour$. The dominant reactions are shown schematically in Fig.\,\ref{figure:ammonia_cycle}. The destruction rates of $\pamm$ and $\oamm$ in reactions with $\hplus$, $\phthree$, and $\ohthree$ are equal, and consequently the o/p-$\amm$ ratio is determined by the nuclear spin branching ratios in the electron recombinations of para-, ortho-, and meta-$\nhfour$. One obtains the relationship
\begin{equation}
{\rm o/p-}\amm = \frac{{\rm m/p-}\nhfour + \frac{1}{3}{\rm o/p-}\nhfour}
                       {1 + \frac{2}{3} {\rm o/p-}\nhfour} \; .
\label{eq:oprule1}
\end{equation}

%

The ammonium ion is predominantly formed in the reactions $\pammplus + \phtwo$ and $\oammplus + \phtwo$. These reactions determine the nuclear spin ratios of $\nhfour$, because the electron recombination rates are equal for the different nuclear spin species. The following relationships are obtained:
\begin{equation}
\begin{array}{lll}
{\rm m/p-}\nhfour &=& \, \frac{5}{8}{\rm o/p-}\ammplus \\
{\rm o/p\,-}\nhfour &=& \frac{15}{8}{\rm o/p-}\ammplus + \frac{3}{2} \, . \\ 
\end{array}
\label{eq:oprule2}
\end{equation}

The primary production pathway to $\ammplus$,  
${\rm NH_2^+} + \htwo \rightarrow \ammplus + {\rm H}$, becomes at late times less important than the charge transfer reaction between $\rm NH_3$ and $\rm H^+$, and in this situation one obtains roughly equal o/p ratios for $\amm$ and $\ammplus$: 
\begin{equation}
{\rm o/p-}\ammplus \sim {\rm o/p-}\amm \; . 
\label{eq:oprule3}
\end{equation} 
Finally, substituting Eqs.\,(\ref{eq:oprule2}) and (\ref{eq:oprule3}) to Eq.\,(\ref{eq:oprule1}), one obtains (with a little algebra) the steady-state ratio
\begin{equation}
{\rm o/p-}\amm \sim 0.4 \; .
\label{eq:oprule4}
\end{equation}
This value is about 10\% lower than the one predicted by our
simulation with the full reaction set. 

The ammonia abundance in interstellar molecular clouds is frequently derived using observations of the (1,1) and (2,2) inversion lines at $\lambda=1.2$\,cm which both represent para-$\amm$. The total ammonia abundance is then derived by assuming ${\rm o/p-}\amm=1$ or that the ortho and para states are populated according to LTE. The latter assumption implies at 10\,K that ${\rm o/p-\amm} = 3.3$. Our result suggests that these previous observational estimates of the total ammonia abundance can be unrealistically large.

Water is produced primarily in the electron recombination of the hydronium ion, $\hthreeo$, and like ammonia, destroyed mainly in the charge transfer reaction with $\hplus$. The recombination of $\phthreeo$ yields both $\ohtwoo$ and $\phtwoo$ at equal ratios, whereas $\ohthreeo$ only yields ortho-water (along with hydroxyl). We obtain the relationship
\begin{equation}
{\rm o/p-}\htwoo = 2\,{\rm o/p-}\hthreeo + 1 \; .
\label{eq:oprule5}
\end{equation}
The production of $\hthreeo$ is dominated by $\htwooplus + \htwo \rightarrow \hthreeo + {\rm H}$. The spin selection rules result in the relationship
\begin{equation}
{\rm o/p-}\hthreeo = \frac{\frac{1}{3} {\rm o/p-}\htwooplus}
                          {1+\frac{2}{3} {\rm o/p-}\htwooplus} \; .
\label{eq:oprule6}
\end{equation}

   \begin{figure}[hbt]
   \centering
   \unitlength=1.0mm
  \begin{picture}(90,65)(0,0)
\color{blue} 

\put(15,10){\circle{10}}
\put(15,10){\makebox(0,0){$\htwooplus$}}

\put(38,35){\circle{10}}
\put(38,35){\makebox(0,0){$\htwoo$}}

\put(52,35){\circle{10}}
\put(52,35){\makebox(0,0){OH}}

\put(45,60){\circle{10}}
\put(45,60){\makebox(0,0){$\hthreeo$}}

\put(75,10){\circle{10}}
\put(75,10){\makebox(0,0){$\ohplus$}}

\put(46,53){\vector(1,-3){4.1}}
\put(40,47){\makebox(0,0){$\el$}}

\put(44,53){\vector(-1,-3){4.1}}

\put(35,30){\vector(-1,-1){15}}
\put(26,25){\makebox(0,0){$\hplus$}}

\put(55,30){\vector(1,-1){15}}
\put(65,25){\makebox(0,0){$\hplus$}}

\put(48,30){\vector(-3,-2){26}}
\put(45,20){\makebox(0,0){$\hthree$}}

\put(15,17){\vector(2,3){25}}
\put(23,36){\makebox(0,0){$\htwo$}}

\put(68,10){\vector(-1,0){46}}
\put(45,5){\makebox(0,0){$\htwo$}}

\end{picture}

  \caption{Dominant reactions involving water and hydroxyl.}
      \label{figure:water_chemistry}
   \end{figure}
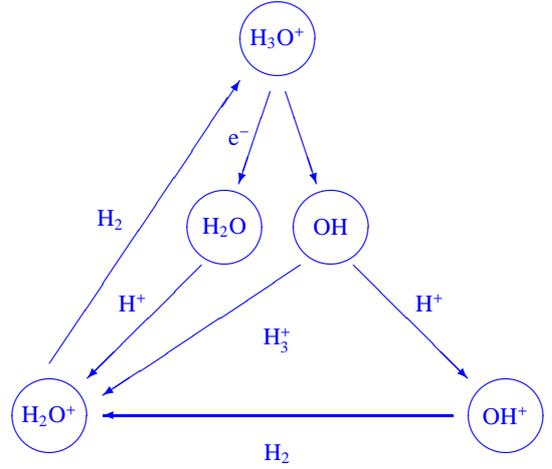
%

The most important reactions determining the abundance of water and related molecules in the gas-phase are shown in Fig.\,\ref{figure:water_chemistry}. The dominant pathway to $\htwooplus$ is the H atom abstraction reaction $\ohplus + \phtwo \rightarrow \mbox{(p or o)}\htwooplus + {\rm H}$, which leads at equal probabilities to $\phtwooplus$ and $\ohtwooplus$. The second reaction of importance is the charge transfer reaction $\htwoo + \hplus$ (with a 20\% share of the $\htwooplus$ production), and the third is proton transfer from $\hthree$ ($\sim10$\%). Omitting the secondary reactions, one obtains
\begin{equation}
{\rm o/p-}\htwooplus \sim 1 \; .
\label{eq:oprule7}
\end{equation}
The full reaction set yields ${\rm o/p-}\htwooplus \sim 1.6$ at late times. The substitution of Eqs.\,(\ref{eq:oprule6}) and (\ref{eq:oprule7}) to Eq.\,(\ref{eq:oprule5}) yields a value  
\begin{equation}
{\rm o/p-}\htwoo \sim 1.4 \; ,
\label{eq:oprule8}
\end{equation}
representing steady state. This is about 20\% lower than the late-time o/p ratio in our simulation. This discrepancy can be understood by the fact that in the very simple model described above, we have omitted the production of $\htwooplus$ from $\htwoo$ and OH.

\onecolumn

\section{New branching ratios for water chemistry \& updates to S13}\label{appendix:water}

\begin{table}[htb]
\caption{The branching ratios of the most important reactions in the water formation network. The reactions given below are the most important in the sense that they are the dominant reactions for the various species, as confirmed by the output by our chemical code. The destruction of $\rm H_3O^+$ by HCN and HNC is important at early times, but electron DR reactions dominate at late times.}
\centering
\begin{tabular}{l l c l l l c}
\hline \hline 
& & Chemical reaction & & & & Branching ratio  \\ \hline
$\rm oH_3^+$ & $\rm O     $ & $\longrightarrow$  & $\rm OH+   $ & $\rm oH_2  $ & & $1.00$ \\ 
$\rm pH_3^+$ & $\rm O     $ & $\longrightarrow$  & $\rm OH+   $ & $\rm pH_2  $ & & $0.50$ \\ 
$\rm pH_3^+$ & $\rm O     $ & $\longrightarrow$  & $\rm OH+   $ & $\rm oH_2  $ & & $0.50$ \\ 
$\rm oH_3^+$ & $\rm O     $ & $\longrightarrow$  & $\rm oH_2O^+$ & $\rm H     $ & & $1.00$ \\ 
$\rm pH_3^+$ & $\rm O     $ & $\longrightarrow$  & $\rm pH_2O^+$ & $\rm H     $ & & $0.50$ \\ 
$\rm pH_3^+$ & $\rm O     $ & $\longrightarrow$  & $\rm oH_2O^+$ & $\rm H     $ & & $0.50$ \\ 
$\rm OH+   $ & $\rm oH_2  $ & $\longrightarrow$  & $\rm pH_2O^+$ & $\rm H     $ & & $0.17$ \\ 
$\rm OH+   $ & $\rm oH_2  $ & $\longrightarrow$  & $\rm oH_2O^+$ & $\rm H     $ & & $0.83$ \\ 
$\rm OH+   $ & $\rm pH_2  $ & $\longrightarrow$  & $\rm pH_2O^+$ & $\rm H     $ & & $0.50$ \\ 
$\rm OH+   $ & $\rm pH_2  $ & $\longrightarrow$  & $\rm oH_2O^+$ & $\rm H     $ & & $0.50$ \\ 
$\rm oH_2O^+$ & $\rm oH_2  $ & $\longrightarrow$  & $\rm oH_3O^+$ & $\rm H     $ & & $0.67$ \\ 
$\rm oH_2O^+$ & $\rm oH_2  $ & $\longrightarrow$  & $\rm pH_3O^+$ & $\rm H     $ & & $0.33$ \\ 
$\rm oH_2O^+$ & $\rm pH_2  $ & $\longrightarrow$  & $\rm oH_3O^+$ & $\rm H     $ & & $0.33$ \\ 
$\rm oH_2O^+$ & $\rm pH_2  $ & $\longrightarrow$  & $\rm pH_3O^+$ & $\rm H     $ & & $0.67$ \\ 
$\rm pH_2O^+$ & $\rm oH_2  $ & $\longrightarrow$  & $\rm oH_3O^+$ & $\rm H     $ & & $0.33$ \\ 
$\rm pH_2O^+$ & $\rm oH_2  $ & $\longrightarrow$  & $\rm pH_3O^+$ & $\rm H     $ & & $0.67$ \\ 
$\rm pH_2O^+$ & $\rm pH_2  $ & $\longrightarrow$  & $\rm pH_3O^+$ & $\rm H     $ & & $1.00$ \\ 
$\rm oH_3^+$ & $\rm oH_2O $ & $\longrightarrow$  & $\rm pH_3O^+$ & $\rm pH_2  $ & & $0.07$ \\ 
$\rm oH_3^+$ & $\rm oH_2O $ & $\longrightarrow$  & $\rm oH_3O^+$ & $\rm pH_2  $ & & $0.08$ \\ 
$\rm oH_3^+$ & $\rm oH_2O $ & $\longrightarrow$  & $\rm oH_3O^+$ & $\rm oH_2  $ & & $0.62$ \\ 
$\rm oH_3^+$ & $\rm oH_2O $ & $\longrightarrow$  & $\rm pH_3O^+$ & $\rm oH_2  $ & & $0.23$ \\ 
$\rm oH_3^+$ & $\rm pH_2O $ & $\longrightarrow$  & $\rm oH_3O^+$ & $\rm pH_2  $ & & $0.25$ \\ 
$\rm oH_3^+$ & $\rm pH_2O $ & $\longrightarrow$  & $\rm oH_3O^+$ & $\rm oH_2  $ & & $0.25$ \\ 
$\rm oH_3^+$ & $\rm pH_2O $ & $\longrightarrow$  & $\rm pH_3O^+$ & $\rm oH_2  $ & & $0.50$ \\ 
$\rm pH_3^+$ & $\rm oH_2O $ & $\longrightarrow$  & $\rm pH_3O^+$ & $\rm pH_2  $ & & $0.13$ \\ 
$\rm pH_3^+$ & $\rm oH_2O $ & $\longrightarrow$  & $\rm oH_3O^+$ & $\rm pH_2  $ & & $0.17$ \\ 
$\rm pH_3^+$ & $\rm oH_2O $ & $\longrightarrow$  & $\rm oH_3O^+$ & $\rm oH_2  $ & & $0.23$ \\ 
$\rm pH_3^+$ & $\rm oH_2O $ & $\longrightarrow$  & $\rm pH_3O^+$ & $\rm oH_2  $ & & $0.47$ \\ 
$\rm pH_3^+$ & $\rm pH_2O $ & $\longrightarrow$  & $\rm pH_3O^+$ & $\rm pH_2  $ & & $0.40$ \\ 
$\rm pH_3^+$ & $\rm pH_2O $ & $\longrightarrow$  & $\rm oH_3O^+$ & $\rm oH_2  $ & & $0.20$ \\ 
$\rm pH_3^+$ & $\rm pH_2O $ & $\longrightarrow$  & $\rm pH_3O^+$ & $\rm oH_2  $ & & $0.40$ \\ 
$\rm oH_3O^+$ & $\rm e^-   $ & $\longrightarrow$  & $\rm OH    $ & $\rm H     $ & $\rm H     $ & $1.00$ \\ 
$\rm pH_3O^+$ & $\rm e^-   $ & $\longrightarrow$  & $\rm OH    $ & $\rm H     $ & $\rm H     $ & $1.00$ \\ 
$\rm oH_3O^+$ & $\rm e^-   $ & $\longrightarrow$  & $\rm oH_2O $ & $\rm H     $ & & $1.00$ \\ 
$\rm pH_3O^+$ & $\rm e^-   $ & $\longrightarrow$  & $\rm pH_2O $ & $\rm H     $ & & $0.50$ \\ 
$\rm pH_3O^+$ & $\rm e^-   $ & $\longrightarrow$  & $\rm oH_2O $ & $\rm H     $ & & $0.50$ \\ 
$\rm oH_3O^+$ & $\rm e^-   $ & $\longrightarrow$  & $\rm OH    $ & $\rm oH_2  $ & & $1.00$ \\ 
$\rm pH_3O^+$ & $\rm e^-   $ & $\longrightarrow$  & $\rm OH    $ & $\rm pH_2  $ & & $0.50$ \\ 
$\rm pH_3O^+$ & $\rm e^-   $ & $\longrightarrow$  & $\rm OH    $ & $\rm oH_2  $ & & $0.50$ \\ 
$\rm oH_3O^+$ & $\rm e^-   $ & $\longrightarrow$  & $\rm oH_2  $ & $\rm H     $ & $\rm O     $ & $1.00$ \\ 
$\rm pH_3O^+$ & $\rm e^-   $ & $\longrightarrow$  & $\rm pH_2  $ & $\rm H     $ & $\rm O     $ & $0.50$ \\ 
$\rm pH_3O^+$ & $\rm e^-   $ & $\longrightarrow$  & $\rm oH_2  $ & $\rm H     $ & $\rm O     $ & $0.50$ \\ 
$\rm oH_3O^+$ & $\rm HNC   $ & $\longrightarrow$  & $\rm HCNH^+ $ & $\rm pH_2O $ & & $0.12$ \\ 
$\rm oH_3O^+$ & $\rm HNC   $ & $\longrightarrow$  & $\rm HCNH^+ $ & $\rm oH_2O $ & & $0.88$ \\ 
$\rm pH_3O^+$ & $\rm HNC   $ & $\longrightarrow$  & $\rm HCNH^+ $ & $\rm pH_2O $ & & $0.38$ \\ 
$\rm pH_3O^+$ & $\rm HNC   $ & $\longrightarrow$  & $\rm HCNH^+ $ & $\rm oH_2O $ & & $0.62$ \\ 
$\rm oH_3O^+$ & $\rm HCN   $ & $\longrightarrow$  & $\rm HCNH^+ $ & $\rm pH_2O $ & & $0.12$ \\ 
$\rm oH_3O^+$ & $\rm HCN   $ & $\longrightarrow$  & $\rm HCNH^+ $ & $\rm oH_2O $ & & $0.88$ \\ 
$\rm pH_3O^+$ & $\rm HCN   $ & $\longrightarrow$  & $\rm HCNH^+ $ & $\rm pH_2O $ & & $0.38$ \\ 
$\rm pH_3O^+$ & $\rm HCN   $ & $\longrightarrow$  & $\rm HCNH^+ $ & $\rm oH_2O $ & & $0.62$ \\ 
\hline
\end{tabular}
\label{tabwater}
\end{table}

\newpage

\begin{table}[htb]
\caption{Rate coefficient data from \citet{LeGal14} included in the present model. We have included some of the Le Gal et al. data ``as is'', while for most reactions we have adopted a total rate coefficient in which contributions from different spin states are added together. When these reactions are separated with our spin-state separation routine, we recover the coefficients given in \citet{LeGal14}.}
\centering
\begin{tabular}{l l c l l l c c c}
\hline \hline 
& & Chemical reaction & & & & $\alpha$ & $\beta$ & $\gamma$  \\ \hline
As is & & & & & & & \\ \hline
$\rm N^+$ & $\rm oH_2$ & $\longrightarrow$  & $\rm NH^+$ & $\rm H$ & & $4.20 \times 10^{-10}$ & $-0.15$ & $44.10$ \\ 
$\rm N^+$ & $\rm pH_2$ & $\longrightarrow$  & $\rm NH^+$ & $\rm H$ & & $8.35 \times 10^{-10}$ & $0.00$ & $168.50$ \\ 
$\rm H^+$ & $\rm oH_2$ & $\longrightarrow$  & $\rm H^+$ & $\rm pH_2$ & & $1.82 \times 10^{-10}$ & $0.13$ & $-0.02$ \\ 
$\rm H^+$ & $\rm pH_2$ & $\longrightarrow$  & $\rm H^+$ & $\rm oH_2$ & & $1.64 \times 10^{-9}$ & $0.13$ & $170.50$ \\ 
$\rm HCO^+$ & $\rm oH_2$ & $\longrightarrow$  & $\rm HCO^+$ & $\rm pH_2$ & & $1.27 \times 10^{-10}$ & $0.00$ & $0.00$ \\ 
$\rm HCO^+$ & $\rm pH_2$ & $\longrightarrow$  & $\rm HCO^+$ & $\rm oH_2$ & & $1.14 \times 10^{-9}$ & $0.00$ & $170.50$ \\ 
$\rm N$ & $\rm OH$ & $\longrightarrow$  & $\rm NO$ & $\rm H$ & & $8.90 \times 10^{-11}$ & $0.20$ & $0.00$ \\ 
$\rm N$ & $\rm NO$ & $\longrightarrow$  & $\rm N_2$ & $\rm O$ & & $7.20 \times 10^{-11}$ & $0.44$ & $12.70$ \\ 
$\rm N$ & $\rm CN$ & $\longrightarrow$  & $\rm N_2$ & $\rm C$ & & $8.80 \times 10^{-11}$ & $0.42$ & $0.00$ \\ 
$\rm N$ & $\rm CH$ & $\longrightarrow$  & $\rm CN$ & $\rm H$ & & $1.70 \times 10^{-10}$ & $0.18$ & $0.00$ \\ 
$\rm C$ & $\rm NO$ & $\longrightarrow$  & $\rm CN$ & $\rm O$ & & $6.00 \times 10^{-11}$ & $-0.16$ & $0.00$ \\ 
$\rm C$ & $\rm NO$ & $\longrightarrow$  & $\rm CO$ & $\rm N$ & & $9.00 \times 10^{-11}$ & $-0.16$ & $0.00$ \\  \hline
Total & & & & & & & \\ \hline
$\rm NH^+$ & $\rm H_2$ & $\longrightarrow$  & $\rm NH_2^+$ & $\rm H$ & & $1.28 \times 10^{-9}$ & $0.00$ & $0.00$ \\ 
$\rm NH_2^+$ & $\rm H_2$ & $\longrightarrow$  & $\rm NH_3^+$ & $\rm H$ & & $2.70 \times 10^{-10}$ & $0.00$ & $0.00$ \\ 
$\rm NH_3^+$ & $\rm H_2$ & $\longrightarrow$  & $\rm NH_4^+$ & $\rm H$ & & $2.40 \times 10^{-12}$ & $0.00$ & $0.00$ \\ 
$\rm H_3^+$ & $\rm NH$ & $\longrightarrow$  & $\rm NH_2^+$ & $\rm H_2$ & & $1.30 \times 10^{-9}$ & $0.00$ & $0.00$ \\ 
$\rm H_3^+$ & $\rm NH_2$ & $\longrightarrow$  & $\rm NH_3^+$ & $\rm H_2$ & & $1.80 \times 10^{-9}$ & $0.00$ & $0.00$ \\ 
$\rm H_3^+$ & $\rm NH_3$ & $\longrightarrow$  & $\rm NH_4^+$ & $\rm H_2$ & & $9.12 \times 10^{-9}$ & $0.00$ & $0.00$ \\ 
$\rm HCO^+$ & $\rm NH$ & $\longrightarrow$  & $\rm NH_2^+$ & $\rm CO$ & & $6.40 \times 10^{-10}$ & $0.00$ & $0.00$ \\ 
$\rm HCO^+$ & $\rm NH_2$ & $\longrightarrow$  & $\rm NH_3^+$ & $\rm CO$ & & $8.90 \times 10^{-10}$ & $0.00$ & $0.00$ \\ 
$\rm HCO^+$ & $\rm NH_3$ & $\longrightarrow$  & $\rm NH_4^+$ & $\rm CO$ & & $1.92 \times 10^{-9}$ & $0.00$ & $0.00$ \\ 
$\rm N_2H^+$ & $\rm NH_3$ & $\longrightarrow$  & $\rm NH_4^+$ & $\rm N_2$ & & $2.30 \times 10^{-9}$ & $0.00$ & $0.00$ \\ 
$\rm N_2H^+$ & $\rm e^-$ & $\longrightarrow$  & $\rm N_2$ & $\rm H$ & & $2.77 \times 10^{-7}$ & $-0.50$ & $0.00$ \\ 
$\rm N_2H^+$ & $\rm e^-$ & $\longrightarrow$  & $\rm NH$ & $\rm N$ & & $2.09 \times 10^{-8}$ & $-0.50$ & $0.00$ \\ 
$\rm NH_2^+$ & $\rm e^-$ & $\longrightarrow$  & $\rm NH$ & $\rm H$ & & $1.17 \times 10^{-7}$ & $-0.50$ & $0.00$ \\ 
$\rm NH_2^+$ & $\rm e^-$ & $\longrightarrow$  & $\rm N$ & $\rm H$ & $\rm H$ & $1.71 \times 10^{-7}$ & $-0.50$ & $0.00$ \\ 
$\rm NH_2^+$ & $\rm e^-$ & $\longrightarrow$  & $\rm N$ & $\rm H_2$ & & $1.20 \times 10^{-8}$ & $-0.50$ & $0.00$ \\
$\rm NH_3^+$ & $\rm e^-$ & $\longrightarrow$  & $\rm NH_2$ & $\rm H$ & & $1.55 \times 10^{-7}$ & $-0.50$ & $0.00$ \\
$\rm NH_3^+$ & $\rm e^-$ & $\longrightarrow$  & $\rm NH$ & $\rm H$ & $\rm H$ & $1.55 \times 10^{-7}$ & $-0.50$ & $0.00$ \\
$\rm NH_4^+$ & $\rm e^-$ & $\longrightarrow$  & $\rm NH_2$ & $\rm H$ & $\rm H$ & $1.22 \times 10^{-7}$ & $-0.60$ & $0.00$ \\
$\rm NH_4^+$ & $\rm e^-$ & $\longrightarrow$  & $\rm NH_2$ & $\rm H_2$ & & $1.88 \times 10^{-8}$ & $-0.60$ & $0.00$ \\
$\rm NH_4^+$ & $\rm e^-$ & $\longrightarrow$  & $\rm NH_3$ & $\rm H$ & & $8.00 \times 10^{-7}$ & $-0.60$ & $0.00$ \\
$\rm H_3O^+$ & $\rm e^-$ & $\longrightarrow$  & $\rm OH$ & $\rm H_2$ & & $3.00 \times 10^{-8}$ & $-0.50$ & $0.00$ \\
$\rm H_3O^+$ & $\rm e^-$ & $\longrightarrow$  & $\rm OH$ & $\rm H$ & $\rm H$ & $2.60 \times 10^{-7}$ & $-0.50$ & $0.00$ \\
$\rm H_3O^+$ & $\rm e^-$ & $\longrightarrow$  & $\rm H_2O$ & $\rm H$ & & $1.10 \times 10^{-7}$ & $-0.50$ & $0.00$ \\
$\rm H_3O^+$ & $\rm e^-$ & $\longrightarrow$  & $\rm H_2$ & $\rm H$ & $\rm O$ & $2.80 \times 10^{-9}$ & $-0.50$ & $0.00$ \\

\hline
\end{tabular}
\label{tabupdates}
\end{table}

\newpage

\section{Calculations at different temperatures}\label{appendix:difftemp}

In this appendix, we present the results of calculations otherwise similar to those presented in Sect.\,\ref{s:results}, but produced assuming either $T_{\rm gas} = T_{\rm dust} = 15$\,K or $T_{\rm gas} = T_{\rm dust} = 20$\,K.

\begin{figure*}[htb]
\centering
\includegraphics[width=16cm]{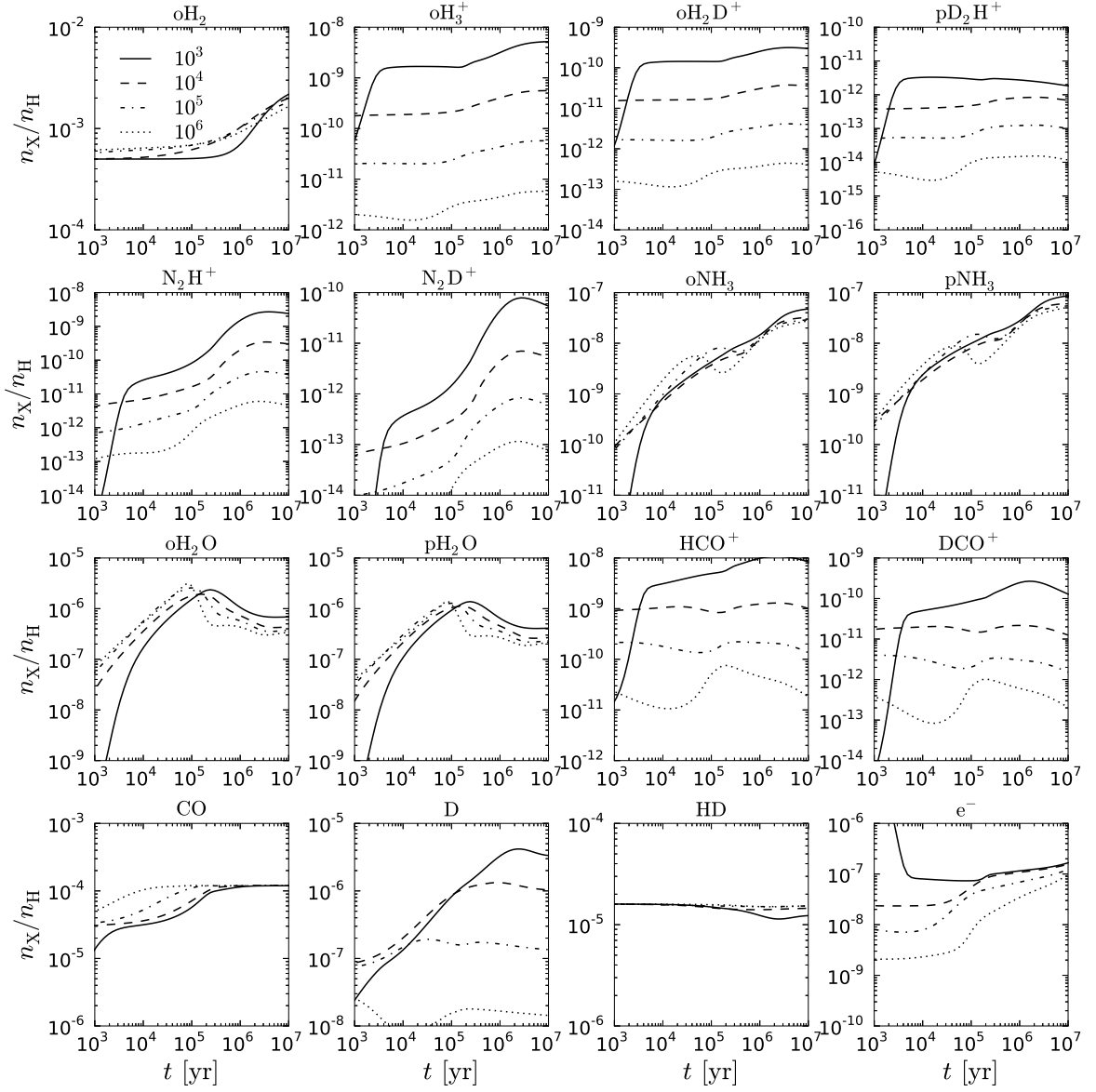}
\caption{As Fig.\,\ref{fig:gasphase10}, but calculated with $T_{\rm gas} = T_{\rm dust} = 15$\,K.
}
\label{fig:gasphase15}
\end{figure*}

\newpage

\begin{figure*}[htb]
\centering
\includegraphics[width=16cm]{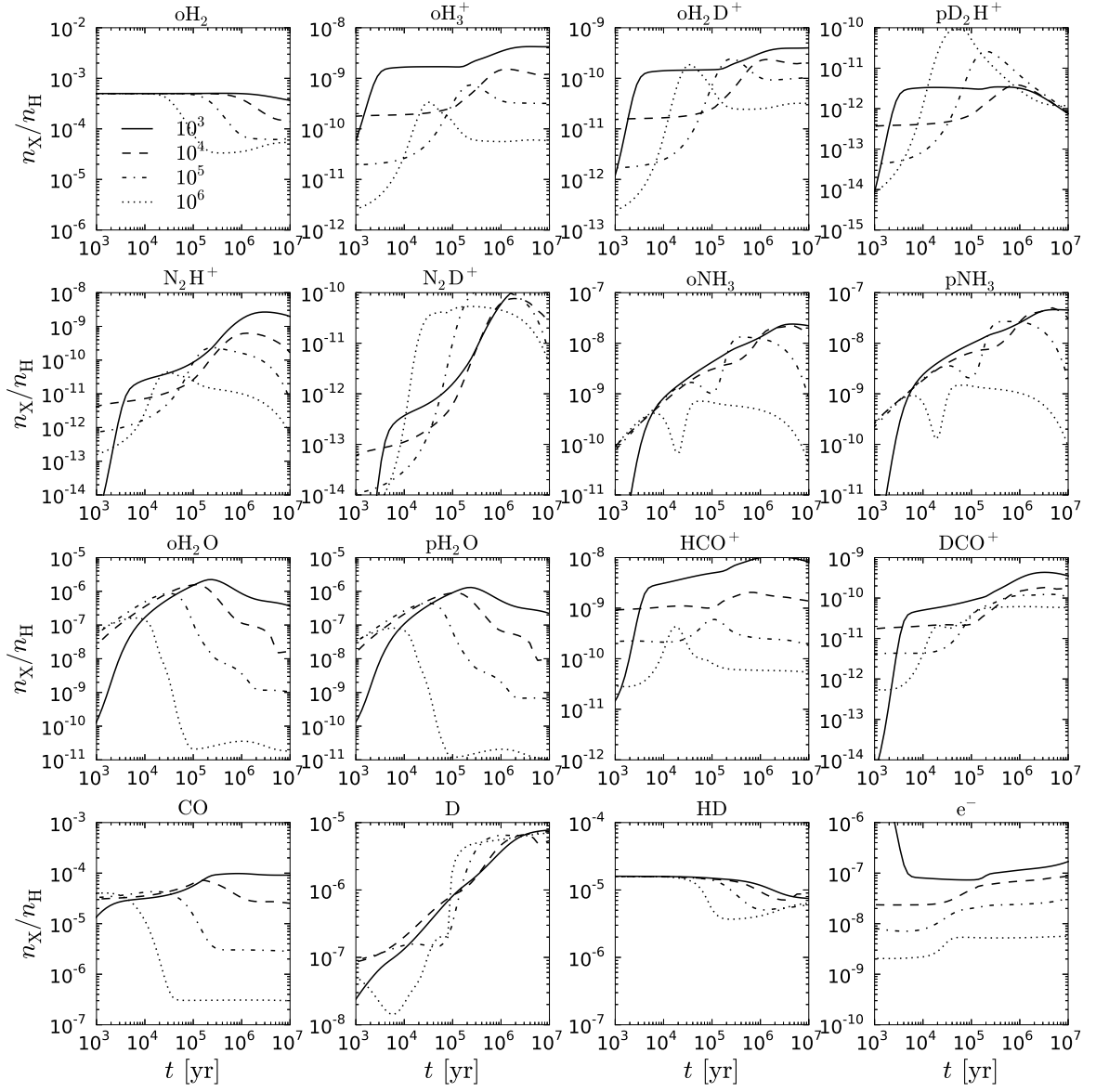}
\caption{As Fig.\,\ref{fig:full10}, but calculated with $T_{\rm gas} = T_{\rm dust} = 15$\,K.
}
\label{fig:full15}
\end{figure*}

\newpage

\begin{figure*}[htb]
\centering
\includegraphics[width=10cm]{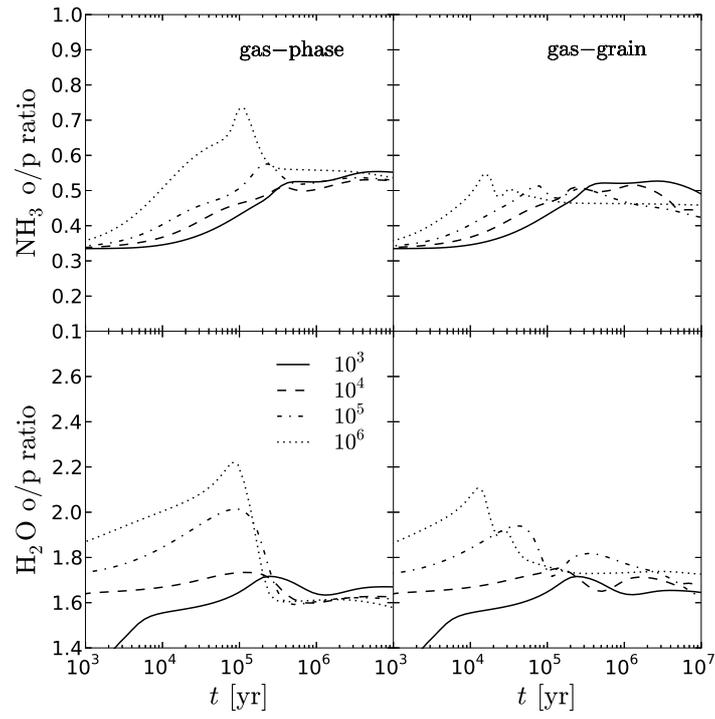}
\caption{As Fig.\,\ref{fig:opratios}, but calculated with $T_{\rm gas} = T_{\rm dust} = 15$\,K.
}
\label{fig:opratios15}
\end{figure*}

\newpage

\begin{figure*}[htb]
\centering
\includegraphics[width=16cm]{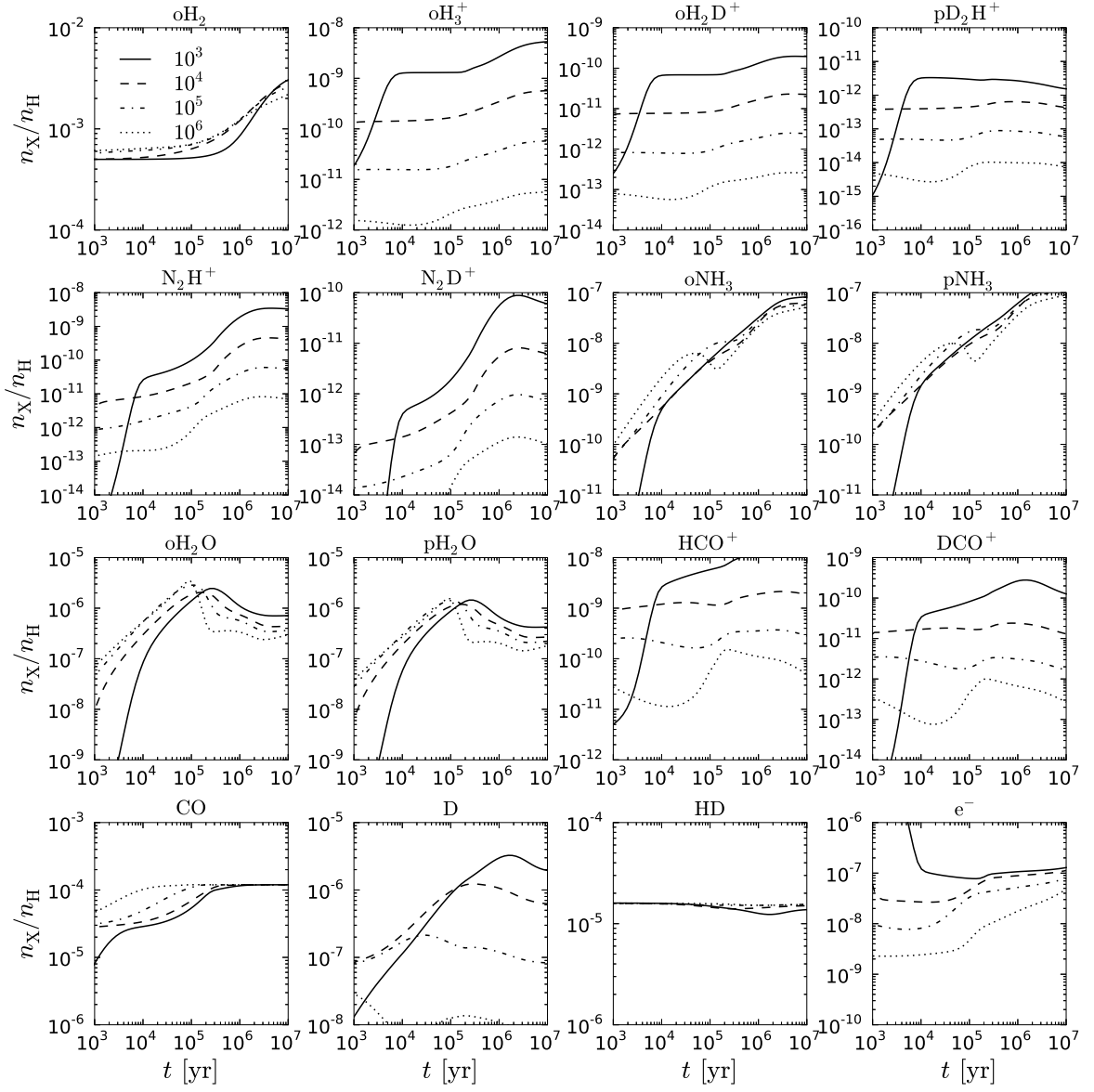}
\caption{As Fig.\,\ref{fig:gasphase10}, but calculated with $T_{\rm gas} = T_{\rm dust} = 20$\,K.
}
\label{fig:gasphase20}
\end{figure*}

\newpage

\begin{figure*}[htb]
\centering
\includegraphics[width=16cm]{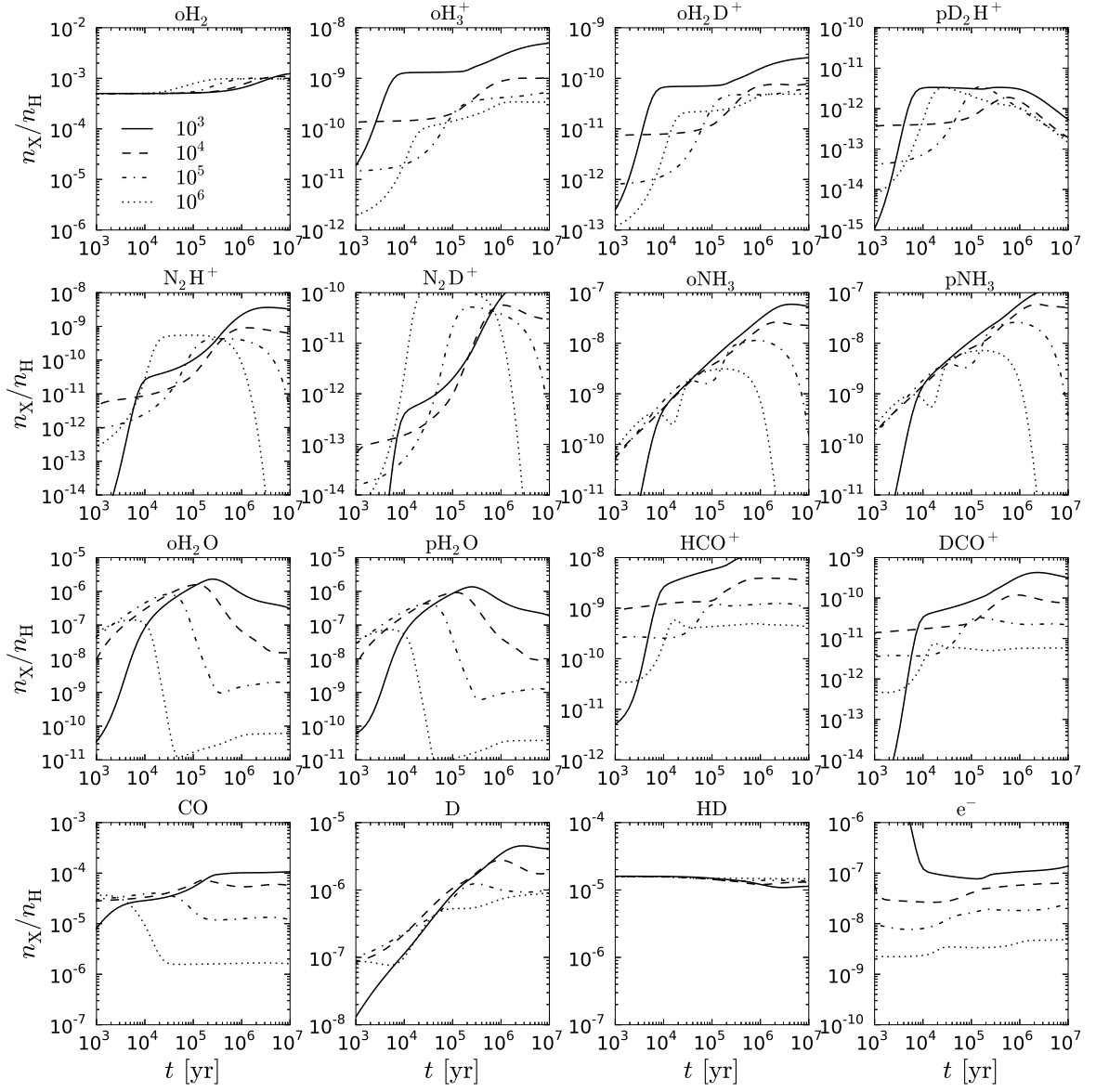}
\caption{As Fig.\,\ref{fig:full10}, but calculated with $T_{\rm gas} = T_{\rm dust} = 20$\,K.
}
\label{fig:full20}
\end{figure*}

\newpage

\begin{figure*}[htb]
\centering
\includegraphics[width=10cm]{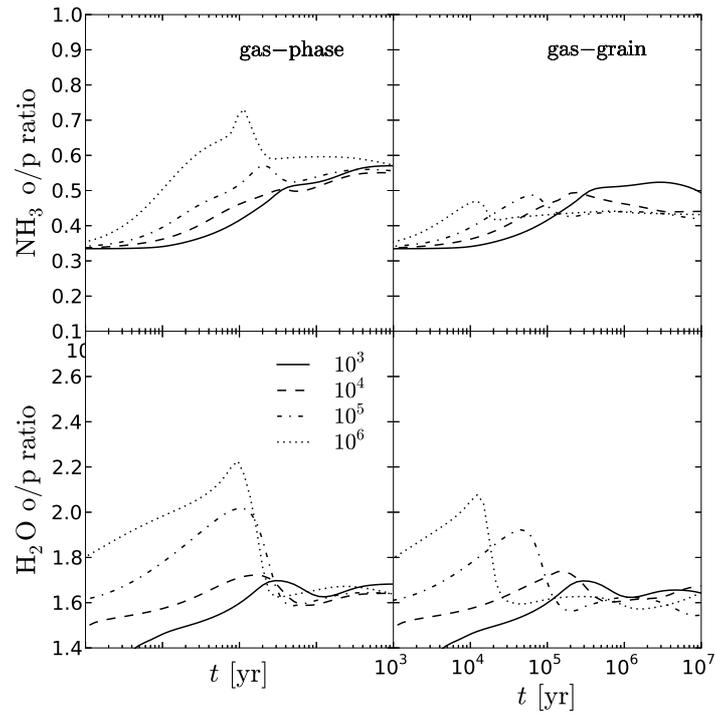}
\caption{As Fig.\,\ref{fig:opratios}, but calculated with $T_{\rm gas} = T_{\rm dust} = 20$\,K.
}
\label{fig:opratios20}
\end{figure*}

\newpage

\section{Calculations without deuterium}\label{appendix:nodeut}

\begin{figure*}[htb]
\centering
\includegraphics[width=17cm]{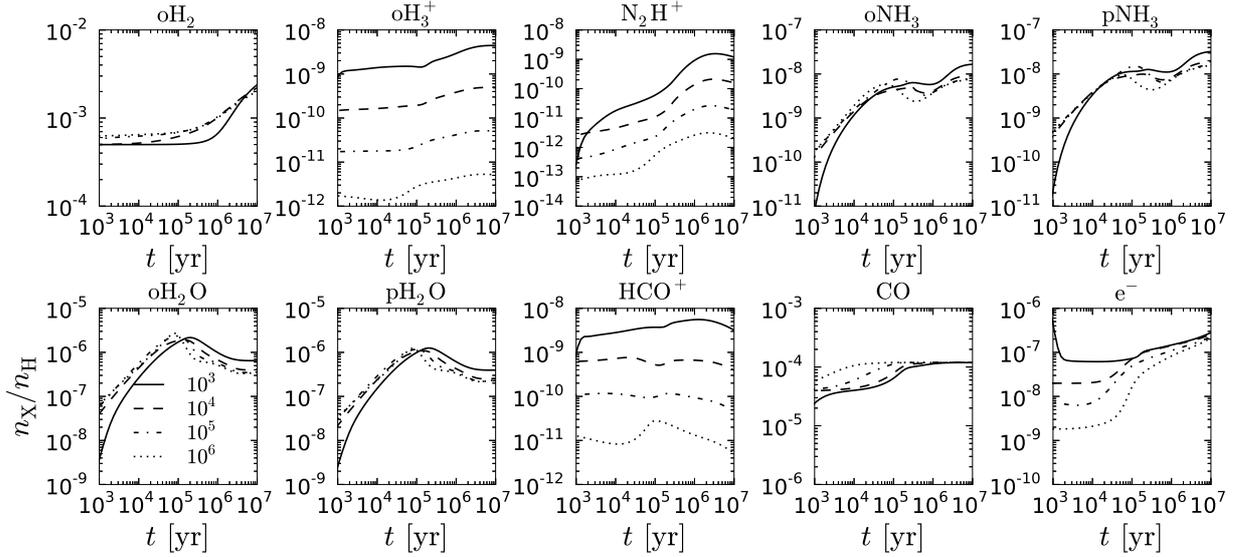}
\caption{Abundances of selected species in a gas-phase model excluding deuterium. This figure should be compared against Fig.\,\ref{fig:gasphase10}.
}
\label{fig:gasphase10_nodeut}
\end{figure*}

\begin{figure*}[htb]
\centering
\includegraphics[width=17cm]{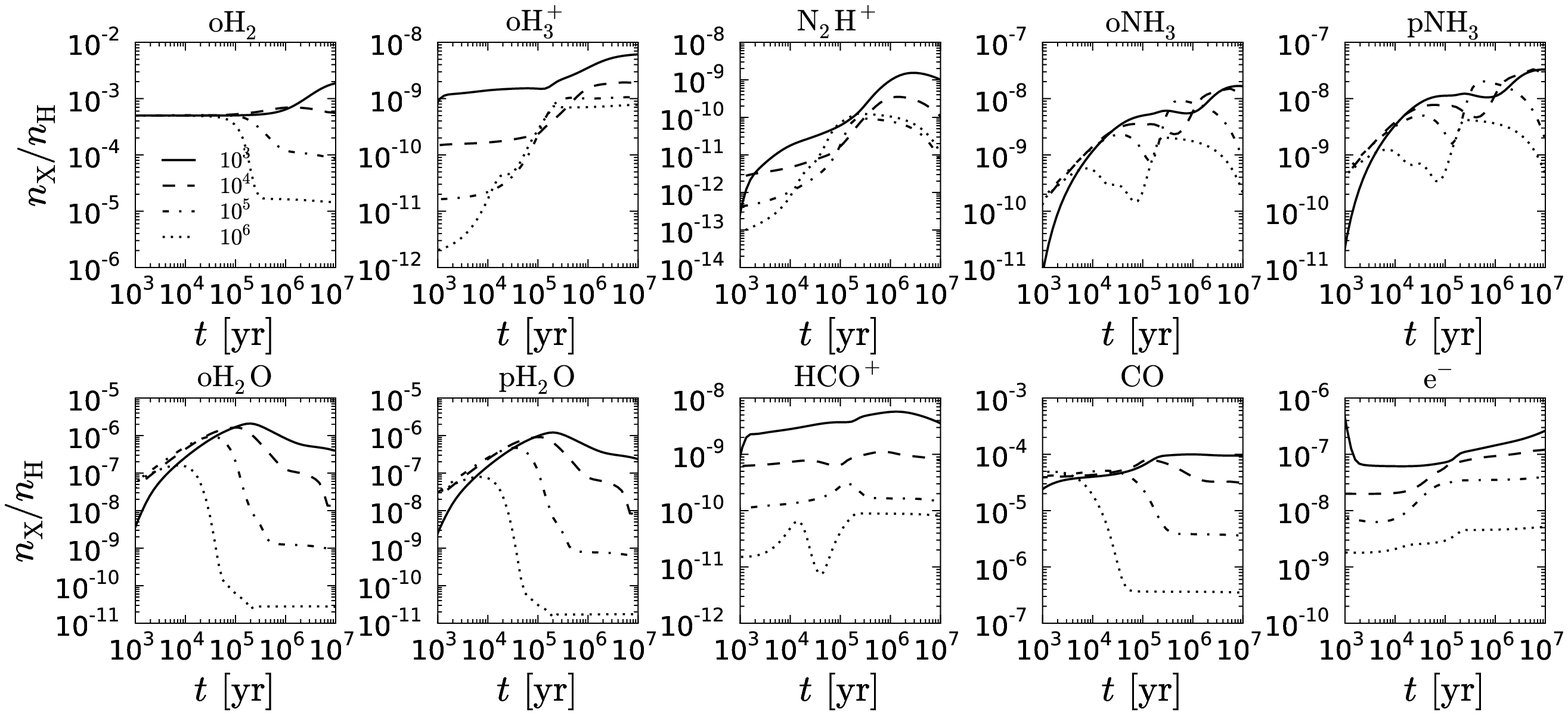}
\caption{Abundances of selected species in a gas-grain model excluding deuterium. This figure should be compared against Fig.\,\ref{fig:full10}.
}
\label{fig:full10_nodeut}
\end{figure*}

\begin{figure*}[htb]
\centering
\includegraphics[width=10cm]{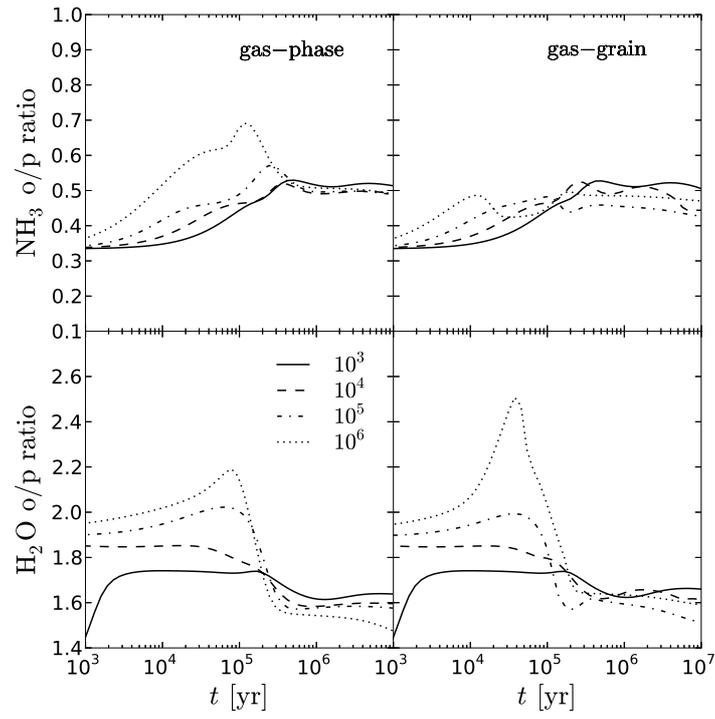}
\caption{As Fig.\,\ref{fig:opratios}, but excluding deuterium.
}
\label{fig:opratios_nodeut}
\end{figure*}

The inclusion of deuterium in a chemical model may decrease the abundances of nondeuterated species especially at high densities in the presence of depletion, where deuterium fractionation is strong. To investigate how our results change when deuterium is excluded, we have run model calculations at $T = 10\,\rm K$ setting the initial HD abundance (i.e., the initial deuterium abundance) to zero. The results of these calculations are presented in Figs.\,\ref{fig:gasphase10_nodeut} to \ref{fig:opratios_nodeut}, which should be compared against Figs.\,\ref{fig:gasphase10}~to~\ref{fig:opratios} in the main text.

It is observed that in the gas-phase model, where there is little deuteration regardless of density, exluding deuterium from the calculations makes virtually no difference in the abundances of the non-deuterated species. Even in the gas-grain model, only very small enhancements are observed in the abundances of non-deuterated species at high density. Also, the o/p ratios of $\rm H_2O$ and $\rm NH_3$ are only very slightly modified in the gas-grain model, and practically not at all in the gas-phase model. Our models thus imply that disregarding deuteration in chemical models when $T \gtrsim 10$\,K will likely not lead to large errors in the abundances of non-deuterated species. We note that we have not carried out a full parameter-space exploration of this issue here, and have only explored the effect of density (and temperature; Appendix\,\ref{appendix:difftemp}) on the results.

\newpage

\newpage

\section{Calculations with quantum tunneling included}\label{appendix:tunneling}

Below, we present the results of calculations performed at $T = 10\,\rm K$, but including quantum tunneling on grain surfaces. When tunneling is included, the thermal diffusion rate defined by Eq.\,(\ref{eq:diff}) is replaced by the tunneling diffusion rate \citep{HHL92}
\begin{equation}\label{eq:Rtunn}
R_i^{\rm diff,q} = {\nu_i \over N_s} \exp \left[-2 \, ( a / \hbar) \, (2 \, m \, k_{\rm B} E_i^{\rm diff})^{1/2} \right] \, ,
\end{equation}
where $a$ is the width of the (rectangular) tunneling barrier. We assume $a = 1\,\AA$. Also, the reaction probability $\kappa_{\ij}$ is in the presence of tunneling replaced by
\begin{equation}\label{eq:ktunn}
\kappa_{ij}^{\rm q} = \exp \left[ -2 \, ( a / \hbar) \, (2 \, \mu \, k_{\rm B} E_a)^{1/2} \right] \, ,
\end{equation}
where $\mu$ is the reduced mass of the reactants (not to be confused with the mean molecular weight of the gas as defined in Eq.\,\ref{eq:grainden}). The reaction rate coefficient assumes the same form as when tunneling is excluded (Eq.\,\ref{eq:ratecoeff}). In the present model, we only allow atomic H and D to tunnel, i.e., Eqs.\,(\ref{eq:Rtunn}) and (\ref{eq:ktunn}) are only used for those reactions where either of these species is present as a reactant.

\begin{figure*}[htb]
\centering
\includegraphics[width=16cm]{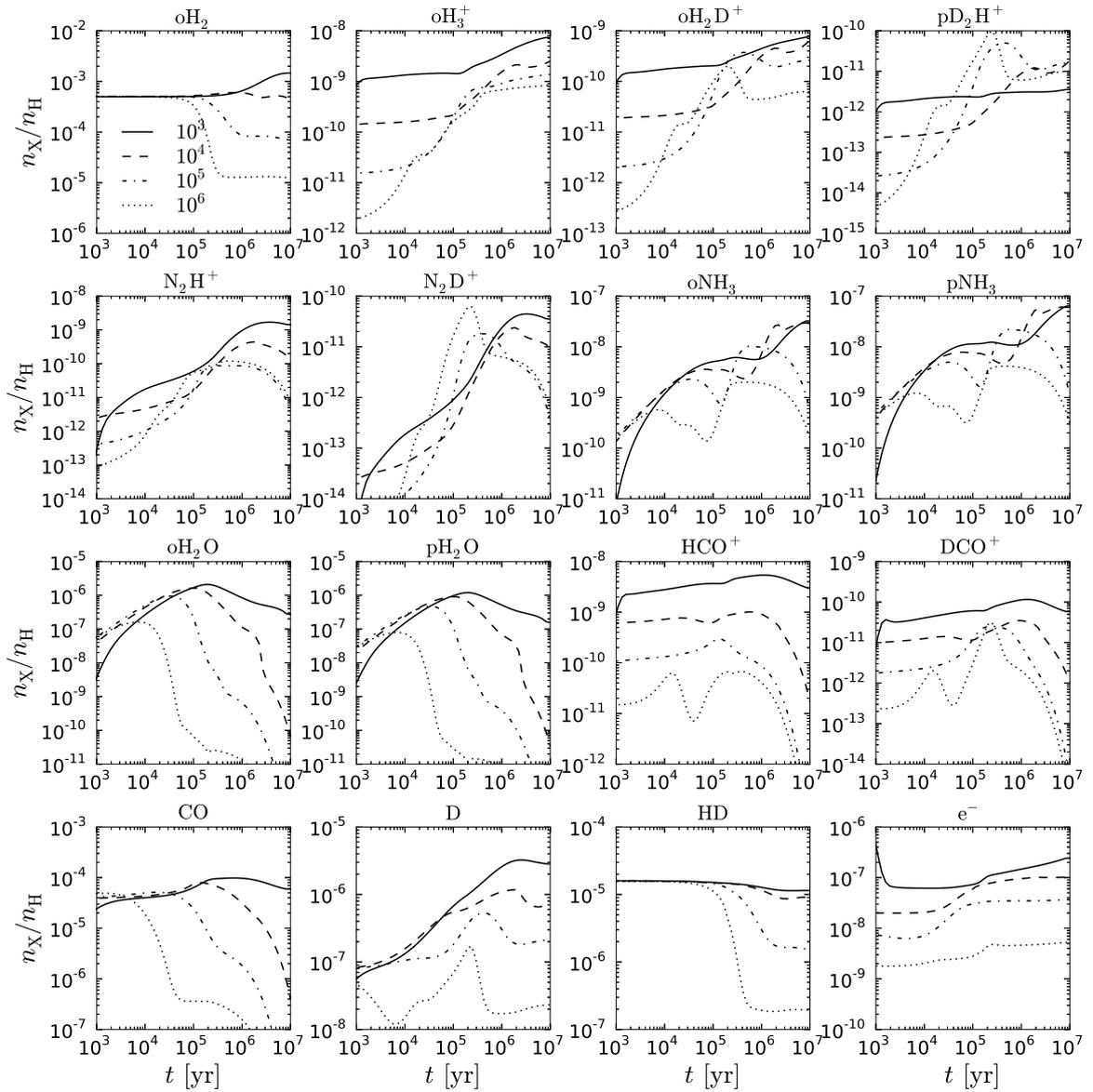}
\caption{As Fig.\,\ref{fig:full10}, but with quantum tunneling included.
}
\label{fig:tunneling10}
\end{figure*}

Figure \ref{fig:tunneling10} presents the results of calculations at $T = 10\,\rm K$ with quantum tunneling included. This figure should be compared against Fig.\,\ref{fig:full10} in the main text. Evidently, tunneling influences our results only at long timescales ($\gtrsim 10^6 \, \rm yr$), and even then the influence on deuteration for example is small. The differences between the models arise because tunneling allows reactions with activation barriers to proceed efficiently. We refer the reader to S13 for more discussion on tunneling and its effects on deuteration.

\end{document}